\RequirePackage{lineno}
\documentclass[superscriptaddress,aps,preprint,amsmath,amssymb,floatfix]{revtex4-1}

\usepackage{graphicx,morefloats}
\usepackage{subfigure} 
\usepackage{color,soul}
\usepackage{caption} 
\usepackage{bm}
\usepackage{amssymb}
\usepackage{amsbsy}
\usepackage{amsmath}
\usepackage{amsthm}
\usepackage{times}
\usepackage{hyperref}
\usepackage{dsfont}
\usepackage{bbm}
\usepackage[normalem]{ulem}
\usepackage{gensymb}

\DeclareMathOperator{\tr}{tr}
\newcommand{\var}{\text{Var}}

\newcommand {\R}{\textcolor {black}}

\newcommand{\be}[0]{\begin{equation}}
\newcommand{\ee}[0]{\end{equation}}
\newcommand{\ba}[0]{\begin{eqnarray}}
\newcommand{\ea}[0]{\end{eqnarray}}
\newcommand{\mx}[0]{\begin{pmatrix}}
\newcommand{\ex}[0]{\end{pmatrix}}

\begin{document}
\hyphenation{va-ni-sh-ing}
\begin{center}
\thispagestyle{empty}

{\large\bf Amplified {multipartite} entanglement witnessed in a quantum critical metal} 
\\
[0.3cm]

Yuan\ Fang$^{1,\dagger}$,
Mounica\ Mahankali$^{1,\dagger}$,
Yiming\ Wang$^{1,\dagger}$,
\\
Lei\ Chen$^{1,\dagger}$,
Haoyu\ Hu$^{2}$,
Silke\ Paschen$^{3}$,
Qimiao\ Si$^{1,\ast}$ 
\\[0.3cm]

$^1$Department of Physics and Astronomy, Extreme Quantum Materials Alliance, Smalley-Curl Institute, Rice University, Houston, Texas 77005, USA\\[-0.cm]

$^2$Donostia International Physics Center, P. Manuel de Lardizabal 4, 20018 Donostia-San Sebastian, Spain\\[-0.cm]

$^3$Institute of Solid State Physics, Vienna University of Technology, Wiedner Hauptstr. 8-10, 1040
Vienna, Austria

\end{center}

\vspace{0.16cm}

{\bf 
Strong correlations in matter promote a landscape of quantum phases and associated quantum critical points. For metallic systems, there is increasing recognition that the quantum criticality goes beyond the Landau framework and, thus, novel means are needed to characterize the quantum critical fluid. Here we do so by studying an entanglement quantity, the quantum Fisher information, in a strange metal system, focusing on the exemplary case of an Anderson/Kondo lattice model near its Kondo destruction quantum critical point. The spin quantum Fisher information peaks at the quantum critical point and indicates a strongly entangled ground state. Our results are supported by the quantum Fisher information extracted from inelastic neutron scattering measurements in heavy fermion metals. Our work elucidates the loss of quasiparticles in strange metals, opens a quantum information avenue to advance the understanding of metallic quantum criticality in a broad range of strongly correlated systems, and points to a novel regime of quantum matter to  realize amplified entanglement.
}
\clearpage
\newpage
\noindent {\bf{Introduction.~~}} Quantum entanglement 
refers to the entwining of particles such
that the quantum state of one cannot be completely described without considering those of the others. This interconnection, lacking a classical counterpart, exhibits unusual properties that defy intuitive understanding~\cite{nielsen2010quantum}.
In condensed matter settings, entanglement can play an important role~\cite{Amico-rmp08}. For example, collective quantum phases of matter in strongly correlated systems, such as the quantum spin liquids and  fractional quantum Hall state, are theoretically expected to have strongly entangled ground states~\cite{wen2004quantum,kitaev2006topological,li2008entanglement}. 
Another class of strongly correlated systems are strange metals, which develop near a quantum critical point (QCP). The theory for quantum criticality in such strongly correlated metals goes beyond the Landau framework of order-parameter fluctuations.
In the case of heavy fermion metals, a critical destruction of Kondo effect~\cite{Si2001,Coleman2001,Senthil2004} has been advanced and extensively evinced~\cite{paschen2021quantum,stefan2020}.
As such, their understanding remains a central question~\cite{paschen2021quantum,Coleman_review,doi:10.1126/science.abh4273,hu2022quantumc}.
Because strange metals are also highly collective, it is natural to ask whether a quantum information perspective will allow for progress in understanding.

\begin{figure}[t!]
    \centering
    \includegraphics[width=0.9\linewidth]{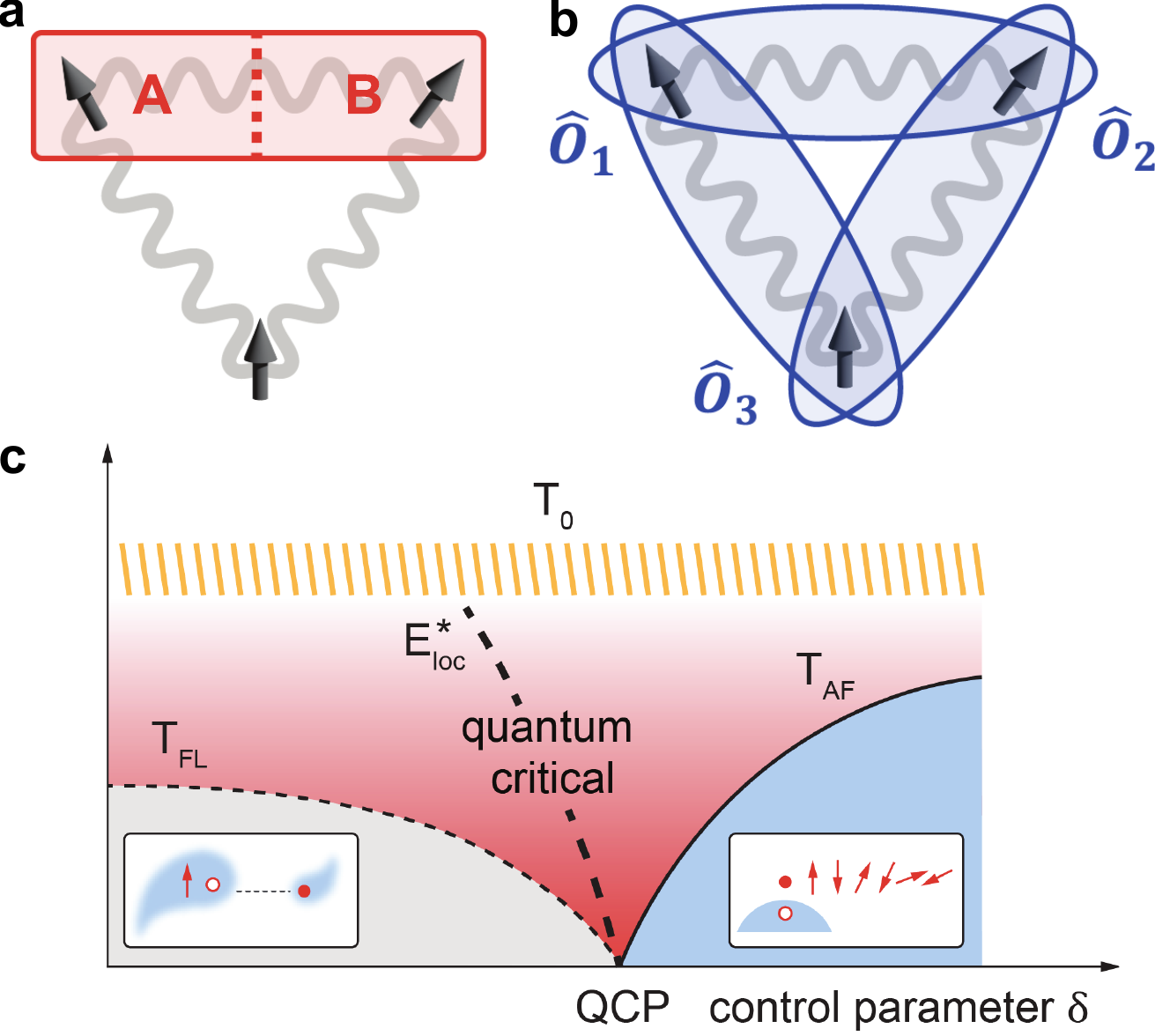}
   \caption{{\bf Illustration of mutual information, quantum Fisher information and Kondo destruction quantum critical point.~~}{\bf a,} Mutual information that detects the entanglement between two subsystems $A$ and $B$. 
    {Here the black arrows indicate spins, the gray wavy lines represent entanglement between two spins, the solid red box means the subsystem of interest and the dashed red line is the separation. }
    {\bf b,} Quantum Fisher information defined for local operators $\widehat{\cal O}_1$, $\widehat{\cal O}_2$, $\widehat{\cal O}_3$ witnesses the multipartite entanglement in the entire system. 
    {Here the blue circles represent the correlation of the local operators for the circled two spins. }
    {\bf c,} Kondo destruction quantum criticality of a Kondo lattice~\cite{hu2022quantumc}. 
    Here the control parameter is the ratio of the RKKY coupling to the bare Kondo temperature, $\delta = I/T_K^0$\,.
    The Kondo destruction energy scale $E_{\text{loc}}^*$ vanishes at the \R{quantum critical point (QCP)}. The three scales, $T_{\text{AF}}$\,, $T_{\text{FL}}$ and $T_0$ correspond to the temperatures of the AF ordering transition, the crossover into the Fermi liquid and the initial onset of Kondo correlations, respectively.
    Cartoons on the two sides of the QCP (the boxes) are adapted from Ref.~\citenum{Prochaska2020}.
    {where the red arrows are $f$ moments and the solid/blank circles represent the particle/hole of $f$ electron. }
    } 
    \label{fig:schematic}
\end{figure}

Here we utilize two quantum information quantities, the mutual information [Fig.\,\ref{fig:schematic}(a)] and quantum Fisher information (QFI) [Fig.\,\ref{fig:schematic}(b)], to analyze the entanglement behavior near a Kondo destruction QCP [Fig.\,\ref{fig:schematic}(c)].
While the entanglement entropy (including the  mutual information) is an effective measure of entanglement, a protocol of how to experimentally probe it in many-body settings has yet to be established, progress in mesoscopic systems notwithstanding~\cite{islam2015measuring,Klich_prl2009}.
By comparison, the QFI is an entanglement witness. Like the Bell's inequality that determines the bipartite entanglement of qubits from correlation functions~\cite{nielsen2010quantum}, the QFI probes the multipartite entanglement in quantum many-body systems~\cite{hyllus2012fisher,Toth2012Multipartite,liu2020quantum,hauke2016measuring,Scheie2024Reconstructing}.
We consider the pertinent theoretical model, the Anderson lattice model (see below), and find the QFI of the spin operator to peak at the QCP, with behavior that indicates a strongly entangled ground state.
Our results elucidate the anomalous quantum dynamics that are often observed in strange metals.

\noindent {\bf{Kondo Destruction Quantum Critical Point.~~}}
The Kondo lattice model --- and its strongly-coupled Anderson lattice counterpart ---
contain two types of degrees of freedom: a lattice of local moments and a band of conduction electrons (see the Methods). The spins of the local moments and conduction electrons are coupled by the Kondo interaction, whereas the local moments are coupled by the Rudermana-Kittel-Kasuya-Yosida (RKKY) interaction~\cite{Si2001,Coleman2001,Senthil2004,Hewson1997}.
Typically, both types of interactions are antiferromagnetic (AF).
The competition between them has been shown~\cite{Si2001,Coleman2001,Senthil2004,hu2022quantumc} to yield quantum phase transitions that involve a heavy Fermi liquid phase and a Kondo destruction phase [Fig.\,\ref{fig:schematic}(c)].
In the heavy Fermi liquid phase, where the Kondo coupling wins over the RKKY interaction, the formation of Kondo singlets between the local spins and conduction electrons gives rise to a large Fermi surface. 
On the other hand, when the RKKY interaction dominates, the correlations among the local spins are detrimental to the formation of any Kondo singlet. As such, at the Kondo destruction QCP where the RKKY and Kondo interactions have the strongest competition, the degree of entanglement among the local spins and conduction electrons represents an intriguing question.
The entanglement entropy has been studied in the past for Kondo systems, for models that involve the local moments at the level of either impurity~\cite{wagner2018long,bayat2012entanglement,bayat2010negativity,alkurtass2016entanglement} or lattice \cite{toldin2019mutual}.
However, the QCP of the Kondo lattice systems has rarely been characterized by any entanglement means~\cite{hu2020quantum}.

\begin{figure}[t!]
    \centering
    \includegraphics[width=0.75\linewidth]{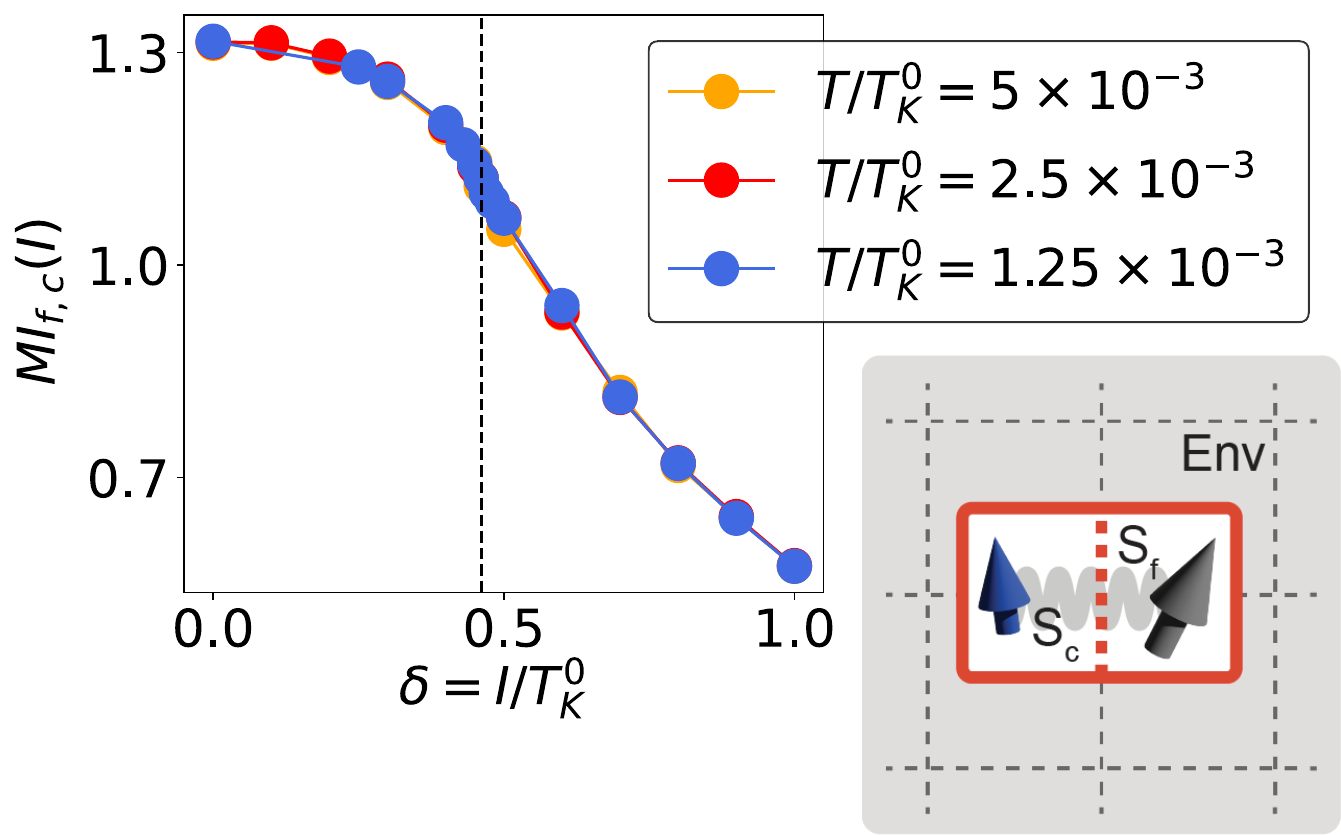}
        \caption{{\bf Mutual information between the local $f$ and $c$ electrons. }Evolution of the mutual information of local $f$ and $c$ electrons with the tuning parameter $\delta = I/T_K^0$. The vertical dashed line indicates the QCP. The schematic diagram of this mutual information is depicted on the right bottom corner,
    where "Env" stands for the rest of the system (the environment).
    {Here, the dashed grids represent the lattice, the red box depicts the impurity subsystem and the red dashed line represents the separation of the $f$ and $c$ electron spins. }
    }
    \label{fig:MI}
\end{figure}

To set the stage for studying the entanglement properties near the QCP, we start from the  mutual information (see the Methods) between the local ($f$) and conduction ($c$) electrons: 
\begin{equation}
    {{\rm MI}_{f,c}} = - S(\hat{\rho}_{f,c}) + S(\hat{\rho}_{f}) + S(\hat{\rho}_c) \, .
\end{equation}
Here, $\hat{\rho}_{f,c}$ denotes the density matrix of the $f$ and $c$ electrons after tracing out the environment (see the schematic in Fig.~\ref{fig:MI}), and $\hat{\rho}_f=\tr_{c} \hat{\rho}_{f,c}$ ($\hat{\rho}_c=\tr_f \hat{\rho}_{f,c}$) is the reduced density matrix of the $f$ ($c$) electrons. 
For definiteness, we consider a square lattice by keeping the nearest neighbor RKKY interaction.
Accordingly, the wavevector dependence of the RKKY interaction has the following form:
$ I_{\bm{q}} = I (\cos q_x + \cos q_y)$,
where $I$ is the strength of the RKKY interaction and $(q_x,q_y)$ is the wavevector at the ordering wavevector $\bm{q} = \bm{Q}=(\pi, \pi)$, 
$I_{\bm{Q}}=-2I$.

In Fig.~\ref{fig:MI}, we show the evolution of the mutual information (${{\rm MI}_{f,c}}$) with respect to the non-thermal parameter ($\delta$) that tunes the phase diagram of the Kondo lattice, namely the ratio of 
the RKKY interaction $I$ to the bare Kondo temperature scale, $T_K^0$.
As this tuning parameter increases, the system undergoes a Kondo destruction quantum phase transition, with the vertical dashed line marking the QCP [{\it cf.} Fig.\,\ref{fig:schematic}(c)]. 
In the Kondo-screened phase, the mutual information is large, essentially saturating the maximal value $2\ln2$ of a spin singlet, which is consistent with the local $f$ moment being strongly bounded to the conduction electrons. The mutual information is monotonically decreased as the RKKY interaction is increased.
Surprisingly, in the Kondo-destroyed side at $\delta >\delta_c$ [{\it cf.} Fig.\,\ref{fig:schematic}(c)], even though the Kondo singlet is destroyed in the ground state, the mutual information ${{\rm MI}_{f,c}}$ remains nonzero.
This result indicates that the Kondo-singlet correlations persist and demonstrates a dynamical Kondo effect~\cite{hu2020quantum}.
Our result sets the stage to probe quantum entanglement when the QCP is approached from {\it both} the Kondo-screened and Kondo-destroyed sides. To do so, we turn to entanglement witness.

\noindent {\bf{Quantum Fisher Information -- General.~~}}
The essence of the mutual information is the entanglement between subsystems A and B.
A different way to detect such entanglement is by measuring the covariance $\tr((\hat{\rho}_{AB} - \hat{\rho}_A\otimes \hat{\rho}_B){\widehat{\cal O}}_A {\widehat{\cal O}}_B)$
where the {Hermitian} operator ${\widehat{\cal O}}_{A/B}$ acts on the A/B subsystem. 
This correlation witnesses the entanglement between the two subsystems with a proper choice of operators, a notion that has only rarely been considered for condensed matter systems.
The QFI corresponds to the summation over all the covariance of site pairs.

Leaving to the Methods and the supplementary information (SI; Sec.\,I) with further details, we note that, on general grounds, the variance is expected to be connected to the correlation functions of the witness operator. Indeed, the QFI of mixed states at temperature $T$ can be determined by the dynamical susceptibility of the operator $\widehat{\cal O}$ (Ref.\,\cite{hauke2016measuring}).
Importantly, through the normalized QFI, the bounds of the QFI provide the entanglement content, which is in the same spirit as Bell's inequality.

\begin{figure}[ht!]
    \centering
    \includegraphics[width=0.55\linewidth]{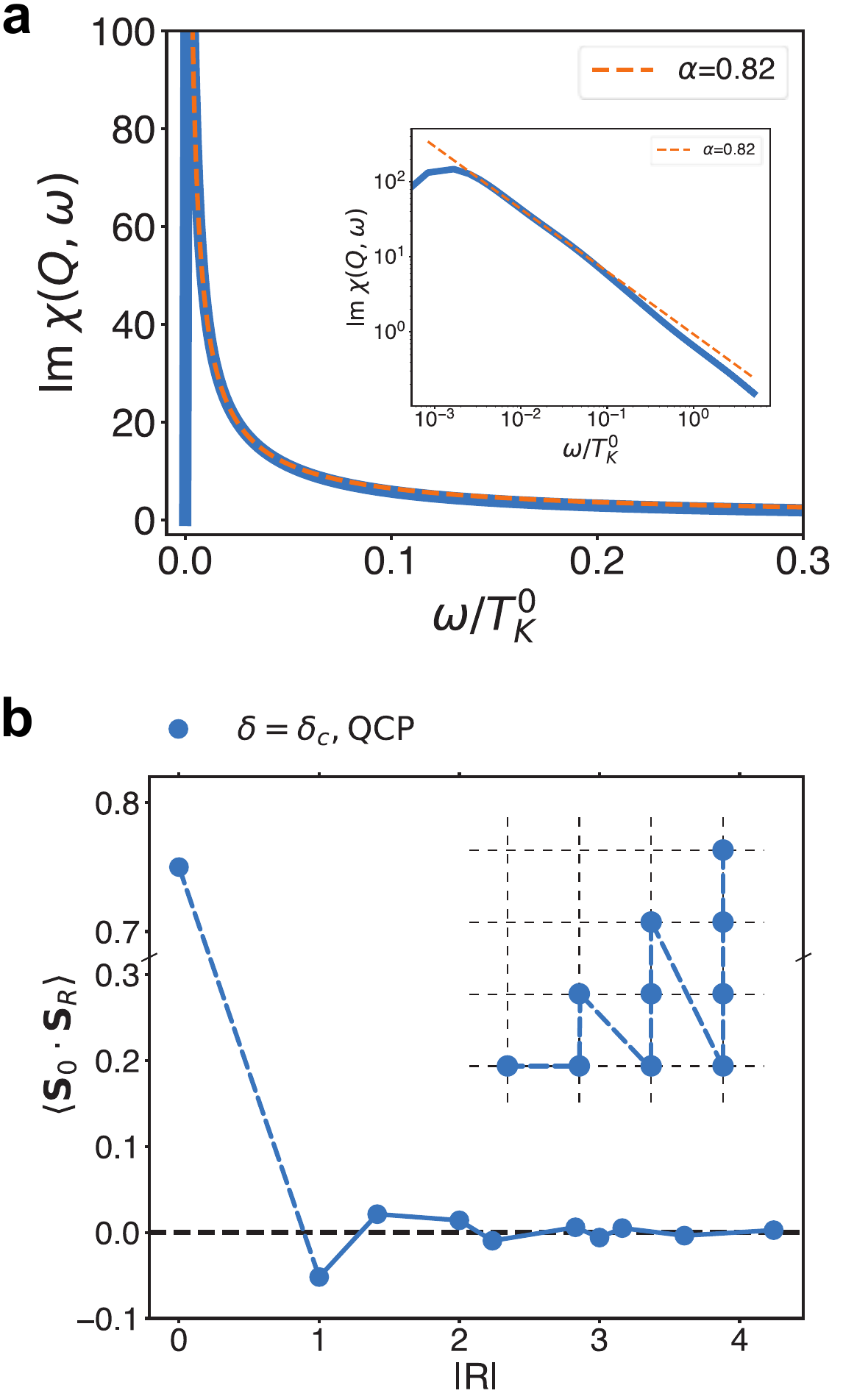}
    \caption{
    {\bf 
    Spin-spin correlation functions at the QCP.~~~}
  (a) The AF dynamical spin susceptibility in real frequency, at the QCP ($\delta = \delta_c$),  is obtained by an analytical continuation from $\chi (\bm{Q}, i\omega_n)$,
  which is simulated by the Monte Carlo method. The blue line denotes the ${\rm Im} \chi(\bm{Q}, \omega)$  at $T/T_K^0=2.5\times 10^{-3}$ obtained from the Pad\'e decomposition, which peaks in the vicinity of $\omega \sim T$. 
  The lower frequency part of the curve follows a power law scaling with critical exponent $\alpha=0.82$.  
  {The fitting in the log-log plot is presented in the inset.}
 (b) The equal-time correlation 
    {function}
    $\langle \mathbf{S}_0\cdot \mathbf{S}_\mathbf{R}\rangle$ at the QCP 
    ($\delta = \delta_c$). Since $|\langle \mathbf{S}_0\cdot \mathbf{S}_\mathbf{R}\rangle| < 0.25$
    for all $\mathbf{R}\neq 0$, the two-tangle is zero:
    there is no pairwise entanglement. 
    The inset displays the lattice sites $\mathbf R$ we select for the calculation.
    }
    \label{fig:S0SR}
\end{figure}

\noindent {\bf{Quantum Fisher Information near the Kondo Destruction Quantum Critical Point.~~}}
We are now in position to present the results of our calculation on the QFI in the Kondo lattice system.
We focus on the spin operator {of the $f$ moments} at the AF wavevector $\bm{Q}$, $\widehat{\cal S}^z=\sum_{i} \widehat{S}^{z}_{i}e^{i\bm{Q}\cdot\bm{R}_{i}}$. 
The normalized QFI density (nQFI) is{~\cite{hauke2016measuring}}
\begin{equation}
    f_{Q} = \frac{2}{\pi} \int_{-\infty}^{\infty} \tanh \frac{ \omega}{2 T} \chi''(\bm{Q}, \omega) d\omega,
\end{equation}
where {we set $\hbar=k_B=1$ (throughout this work) and} $\chi''(\bm{Q}, \omega)$ is the imaginary part of the dynamical spin susceptibility at wave vector $\bm{Q}$. 
[Note that $\chi''(\bm{Q},\omega)=\chi''_{{\cal S}^z}(\omega)/N$, with $N$ the total number of sites.]
In practice, the Monte Carlo simulation of the equations in the extended dynamical mean field theory (EDMFT) (see the Methods) is performed on the imaginary frequency axis.
We carry out an analytical continuation to obtain the imaginary part of the dynamical spin susceptibility in real frequency, $\chi '' (\bm{Q}, \omega)$ (for details, see the SI, Sec.\,III). 
{Fig.~\ref{fig:S0SR}(a) shows} $\chi '' (\bm{Q}, \omega)$ at the QCP, which is obtained by a Pad\'e decomposition. The lower-frequency part of the curve follows a power law scaling behavior with an anomalous critical exponent $\alpha \approx 0.82$; its fractional nature is a salient feature of the Kondo destruction QCP \cite{Si2001}.

\begin{figure}[t!]
    \centering
    \includegraphics[width=\linewidth]{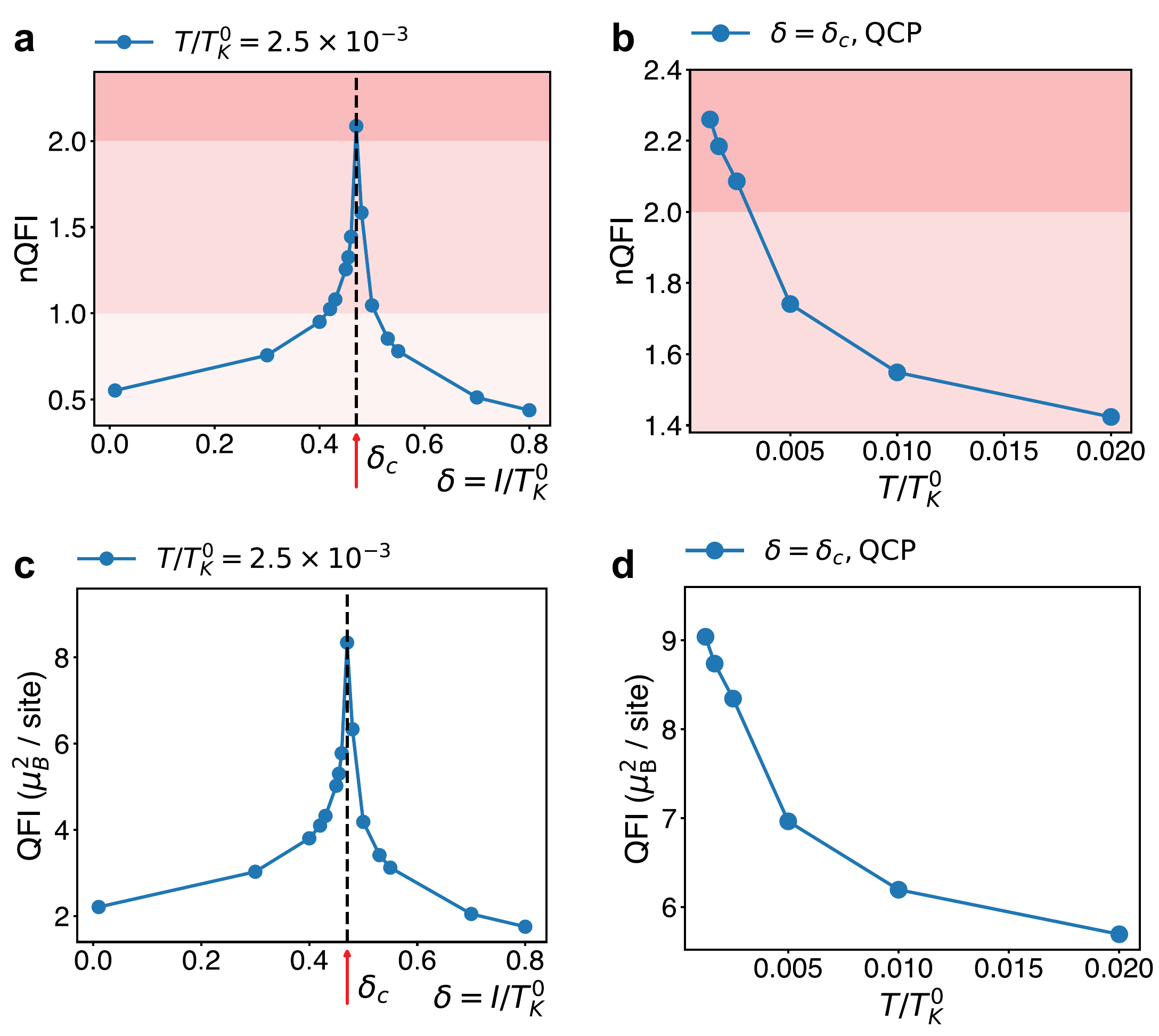}
    \caption{{\bf 
    The  Quantum Fisher information density.~~}
    (a) The low-temperature normalized QFI density 
    (nQFI) of the AF spin operator, $f_Q$, as a function of the tuning parameter $\delta=I/T_K^0$. 
    The vertical dashed line and the red arrow indicate the location of the QCP. The nQFI is peaked at the QCP, 
    where it exceeds $2$,
    indicating that the ground state contains at least three-partite entanglement.
    The dark (light) red shading, here and in panel (b), marks the regime where the system is at least $3$- ($2$-) partite entangled.
    (b) The 
    nQFI vs. temperature $T$ (normalized by the bare Kondo temperature, $T_K^0$) at the quantum critical coupling, $\delta_c=I_c/T_K^0$.
    The
    results are obtained in the quantum critical regime with $\delta_c=0.47$, except for the lowest temperature $T=1.25 \times 10^{-3}$, which requires a finer tuning to reach quantum criticality and is calculated at $\delta_c=0.465$.
    (c) (d) The un-normalized QFI density of the AF magnetization operator, 
    $f_Q(g\mu_B)^2$, 
    suitable for comparison with its inelastic neutron scattering determination. Here, $g$ is taken to be $2$.
    }
    \label{fig:QFI}
\end{figure}

We show the normalized QFI density, at a very low temperature ($T=2.5 \times 10^{-3} T_K^0$), versus the tuning parameter $\delta=I/T_K^0$ in Fig.~\ref{fig:QFI}(a), where the vertical dashed line marks the QCP. The left and right hand sides of the QCP correspond to the Kondo screened phase and Kondo destroyed phase [Fig.\,\ref{fig:schematic}(c)], respectively. 
We find $f_{Q}$ to display a sharp peak at the QCP, reaching the value around $2.2$. This peak exceeds the bound of $2$, implying that the ground state contains at least three-partite entanglement according to Eq.~(\ref{eqn:upperbound}).

We next fix the system to be at $\delta = \delta_c$ and change the temperature (with the QCP located at $T=0)$. 
As shown in Fig.~\ref{fig:QFI}(b), the nQFI monotonically increases with decreasing temperature and reaches the same low-temperature regime of at least three-partite entanglement.

To probe further the entanglement, we compute the two-tangle of spin pairs, which is a measure of the entanglement between 
two spins~\cite{PhysRevLett.80.2245} (see the SI, Sec.~II).
At the QCP where the spin parity and translation symmetry are preserved and the magnetic order vanishes, the two-tangle is expressed as $\tau_{0\mathbf{R}} = |\max \{ 0,-\frac12\pm 2\langle \mathbf{S}_0\cdot \mathbf{S}_\mathbf{R}\rangle \} |^2 $
(Refs.\,\citenum{amico2004dynamics,scheie2023proximate}), in which the correlation function $\langle \mathbf{S}_0\cdot \mathbf{S}_\mathbf{R}\rangle = \frac{1}{2\pi N_{\bm{q}}}\sum_{\bm{q}} e^{-i\bm{q} \cdot \mathbf{R} }\int d\omega \chi(\bm{q}, \omega)$. 
Here $N_{\bm q}$ is the number of momentum points in the Brillouin zone and $\chi(\R{\mathbf q}, \omega)$ is the dynamical susceptibility of the spin operator.

We present this correlation function in Fig.~\ref{fig:S0SR}(b). It is evident that $\tau_{0\mathbf{R}}=0$ for every pair of spins. 
Since our QFI reveals multipartite entanglement in the system, the vanishing two-tangles mean that the entanglement is distributed among multiple spins rather than being confined to pairwise spins. It manifests the quantum monogamy and the Coffman-Kundu-Wootters inequality~\cite{coffman2000distributed,PhysRevLett.96.220503}. For example, a maximally entangled state of spins with entanglement distributed among all spins have no entanglement inside any subspace (see the SI, Sec.\,II; in particular, Fig.\,S1).
The combination of a multipartite QFI and vanishing two-tangle provide evidence for strong entanglement in the system.

Finally, to facilitate comparison with inelastic neutron scattering experiments, we show in Fig.~\ref{fig:QFI}(c)(d) the (un-normalized) QFI density appropriate for this spectroscopy, {\it viz.} for the AF magnetization operator,
$g\mu_{\rm B} \sum_{i} \widehat{S}^{z}_{i}e^{i\bm{Q}\cdot\bm{R}_{i}}$.
Here, for an order-of-magnitude estimate of the QFI, we have taken the $g$-factor to be $2$, which is generally considered to be suitable for the 4$f$ crystal field ground state in heavy fermion systems.

\noindent {\bf{{Discussion and Comparison with Experiment.}~~}}
Several points are in order.
First of all, our theoretical results at the critical coupling can be tested by experiments.
Both the temperature dependence and magnitude of the QFI
are supported by the spin QFI we have extracted [see the SI, Sec.\,IV; {\it cf.} Fig.\,\ref{fig:S2}(a), compared with Fig.\,\ref{fig:QFI}(d)] from the inelastic neutron scattering data~\cite{schroder2000onset} of CeCu$_{5.9}$Au$_{0.1}$; this is a canonical heavy fermion metal in which the Kondo effect is associated with localized spins, as in our model, and it hosts a QCP that has been recognized to be of the Kondo destruction type. Another compound to consider is Ce$_3$Pd$_{20}$Si$_6$:
Even though the local degrees of freedom here are more complex compared to our model, involving entwined spin and orbital, the QFI result recently determined by inelastic neutron scattering spectroscopy at a field-induced QCP also compares well with our theory [{\it cf.} Fig.\,3 of Ref.~\citenum{Paschen2024}, compared  with Fig.\,\ref{fig:QFI}(d)]. A further support is that the measured spin dynamics in both materials are singular and have a fractional critical dynamical exponent, as in our theoretical result; the measured exponents are comparable to the theoretical value of $\alpha=0.82$. In other condensed matter contexts, the QFI has only been measured in several insulating quantum magnets. The low-temperature value for the normalized QFI density we have calculated for the Kondo destruction QCP rises to the values determined in candidate materials for such highly entangled ground states as two-dimensional spin liquids{~\cite{George2020Experimental,Pratt2022Spin}} ($3.4$ for $\rm KYbSe_2$, a layered system that realizes a spin-$1/2$ Heisenberg model on triangle lattice, Ref.\,\citenum{scheie2023proximate}) and one-dimensional spin liquid with fractionalized spin excitations ($3.8$ for KCuF$_3$, a quasi-one-dimensional system that  realizes a spin-$1/2$ Heisenberg chain, Ref.\,\citenum{PhysRevB.103.224434,*PhysRevB.107.059902}), respectively.
We note that, in our EDMFT calculation, the momentum dependence of
the QFI  is captured through Eq.~(\ref{eqn:chi_latt}).

Secondly, we find the QFI to be peaked at the QCP. This result, testable by future experiments on quantum critical heavy fermion systems, establishes the strange metal regime as amplifying quantum entanglement. It is to be contrasted with the observation in certain insulating quantum magnet~\cite{PhysRevLett.127.037201}. Equally important, we find that the multipartite entanglement is witnessed only in the immediate vicinity of the QCP: it no longer is the case when the tuning parameter $\delta$ moves outside of the quantum critical regime [Fig.~\ref{fig:QFI}(a)] and the system goes back to the Fermi liquid regime described in terms of quasiparticles [{\it cf.} Fig.~\ref{fig:schematic}(c)]. Multipartite entanglement is associated with unusual quantum many-body dynamics~\cite{Brenes2020,JSmith2016}.
As such, our work suggests that, in strange metals, multipartite entanglement provides a general characterization of their anomalous dynamics.

More specifically, our work elucidates the key characteristic of strange metallicity, { viz.} the loss of Landau quasiparticles \cite{hu2022quantumc,Liyang-Chen2023}.
Our mutual information calculation demonstrates a nearly saturated Kondo entanglement in the paramagnetic (heavy fermion) phase, which describes the formation of Kondo singlets in the ground state 
{[cf. the Supplementary Note 5 and Supplementary Fig.~4{\bf a}]. }
This implicates a $1$-to-$1$ correspondence: each local moment is converted into a heavy quasiparticle in the excitation spectrum 
{[cf. the Supplementary Fig.~4{\bf c}, the right part], }
through the process of breaking up the Kondo singlet 
{[cf. the Supplementary Fig.~4{\bf c}, the left part]. That
the calculated mutual information for the Fermi liquid phases
leads to pictures consistent with previous intuitive understandings  
sets the stage for new understandings from the QFI calculation, especially for the strange metal in the quantum critical regime.}

The result from our QFI calculation is the development of multipartite entanglement among the $f$-moments 
{in the quantum critical regime}. 
It describes a strong entanglement among the local spins
of  the resonant-valence-bond (RVB) type
{(cf. Supplementary Fig.~4{\bf d},\,{\bf e})}, which, through the entanglement monogamy, characterizes the
suppression of the $f$-$c$ (Kondo) entanglement and, by extension, the strange metal's loss of quasiparticles.
This represents a central  insight into the inner workings of strange metallicity 
that is enabled by our study of the
quantum Fisher information. Further discussion of this point can be found in
{Supplementary Note 5.}

Our elucidation of the strange metallicity from the entanglement perspective connects well with the existing phenomenology.
For example, the QCP of the canonical heavy fermion strange metal YbRh$_2$Si$_2$ is magnetic (AF, to be precise) in nature, 
involving a zero-temperature transition between AF and paramagnetic metallic phases.
Yet, its charge response is found to be critical~\cite{Prochaska2020}, a property that also arises in model calculations~\cite{Cai_prl2020}.
The present study implicates an enhanced quantum entanglement as underlying the phenomenon. 
In turn, our work suggests future studies to probe the QFI of other degrees of freedom that are enabled by the strange metallicity.
For example, it is opportune to study the charge QFI from the singular charge fluctuations observed in a cuprate strange metal near optimal superconductivity~\cite{doi:10.1073/pnas.1721495115}.
Recent proposals for further QFI spectroscopies concern the resonant inelastic X-ray scattering (RIXS)~{\cite{hales2023witnessing,Baykusheva2023Witnessing}} and angle-resolved photoemission spectroscopy (ARPES)~\cite{malla2023detecting}. Thus, there is considerable prospect for further studies of multipartite entanglement in strange metals.

\noindent {\bf{Summary.}~~~}
We have carried out the first theoretical study of entanglement witness in a model for strange metallicity and beyond-Landau quantum criticality.
Our results provide much-needed characterization of beyond-Landau quantum critical metals.
The quantum Fisher information reveals amplified entanglement at the quantum critical point.
The witnessed multipartite entanglement brings out new insights into strange metals' anomalous dynamics and loss of quasiparticles.
Our work showcases a new window into the quantum correlations that underlie a wide range of strongly interacting metallic systems, and points to broad classes of correlated gapless quantum matter with strange metallicity as a promising setting to witness enhanced entanglement.

\medskip

\noindent{\bf\large Methods}
\\
\noindent {\bf{Quantum criticality in the Anderson lattice model.~~} }
We consider the $\mathrm{SU(2)}$ 
periodic Anderson lattice model, which takes the form:
\begin{equation}
\label{eq:AM}
\begin{aligned}
    H &= \frac{U}{2} \sum_{i} \left[ \sum_{\sigma} f_{i,\sigma}^{\dagger}f_{i,\sigma} -1\right]^2 + \sum_{ij} I_{ij} \bm{S}_{i} \cdot \bm{S}_j \\
    & + V\sum_{i,\sigma}\left( c_{i,\sigma}^{\dagger} f_{i,\sigma} + f_{i\sigma}^{\dagger}c_{i,\sigma} \right) + \sum_{p,\sigma} \epsilon_{p} c^{\dagger}_{p,\sigma} c_{p,\sigma} \,.
\end{aligned}
\end{equation}
Here, $f_{i\sigma}^{\dagger}$ ($c_{i\sigma}^{\dagger}$) creates a local (conduction) electron with spin $\sigma$ at site $i$. The hybridization $V$ couples the local $f$-electron with the conduction $c$-electron, which  has a dispersion $\epsilon_{p}$. 
Moreover, the on-site Coulomb repulsion is represented by $U$. When $U$ is sufficiently large, the model is equivalent to the Kondo lattice Hamiltonian, with a Kondo coupling $J_K\sim \frac{4V^{2}}{U}$. 
Finally, $\bm{S}_{i} = f^{\dagger} \frac{\bm{\sigma}}{2} f$ represents the spin operator of the local moment, and $I_{ij}$ describes the  AF RKKY interaction between the local moments.  
We consider an RKKY interaction $I_{ij}$. Its Fourier transformation, $I_{\bm{q}}$, reaches the most negative value at the AF wave vector $\bm{Q}$, with $I_{\bm{Q}}=-2I$.

We treat the Hamiltonian described in Eq.~(\ref{eq:AM}) by the extended dynamical mean-field (EDMFT) method, within which the dynamical competition between the Kondo hybridization and RKKY interaction is taken into account appropriately~\cite{hu2022extended,Si1996,Smith2000,Chitra2000}.
Through EDMFT, the correlation functions of the lattice model are calculated in terms of those of a self-consistent Bose-Fermi Anderson (BFA) model, in which the local $f$ electron couples with a fermionic bath and a bosonic bath. The fermioninic bath comes from the hybridization with the conduction electrons and the bosonic bath describes the spin fluctuations from the RKKY interaction. The action after integrating out both baths takes the form:
\begin{equation}
\begin{aligned}
    S_{BFA} & = \int_{0}^{\beta} d\tau \left[ \sum_{\sigma} f_{\sigma}^{\dagger} \partial_{\tau} f_{\sigma} +{\eta_{\text{loc}}} S^{z} +\frac{U}{2} \left(n_f -1\right)^2 \right] \\
    &  - \int_{0}^{\beta} d\tau d\tau' \bigg[ \sum_{\sigma} f_{\sigma}^{\dagger}(\tau) V^2 \mathcal{G}_{c}(\tau-\tau') f_{\sigma} \\
    & + \frac{1}{2} \sum_{a} S^{a}[\chi_{0}^{-1}(\tau-\tau')] S^{a}(\tau') \bigg] \, ,
\end{aligned}
\end{equation}
where $\beta=1/T$ and $n_{f} = \sum_{\sigma} f_{\sigma}^{\dagger}f_{\sigma}$. 
{Note, here $\tau$ stands for the imaginary time.}
{(Again, we set $\hbar=k_B=1$ throughout this work)}. The static Weiss field ${\eta_{\text{loc}}}$ is introduced to capture the AF order, while $\mathcal{G}_{c}$ and $\chi_{0}$ denote the Green's function of the fermionic and bosonic bath, respectively. 
Note that $a\in\{x, y, z\}$. 
The self-consistent conditions are:
\begin{equation}
\begin{aligned}
    \chi_{\text{loc}}^{a}(i\omega_n) &= \int d\epsilon \frac{\rho_{I}(\epsilon)}{\epsilon +M^{a}(i\omega_n)} \\
    M^{a}(i\omega_n) &= 1/\chi_0^{a}(i\omega_n) +
    1/\chi_{\text{loc}}^{a}(i\omega_n) \\
    \mathcal{G}_c(i\omega_n) &=\int d\epsilon \frac{\rho_0(\epsilon) }{-i\omega_n + \epsilon + \Sigma_{c}(i\omega_n)} \\
    {\eta_{\text{loc}}} &= -[2I- \chi_0^{a}(i\omega_n=0)]m_{AF}\,.
\end{aligned}
\end{equation}
Here, $\Sigma_{c}$ is the conduction-electron self-energy and $M^{a}$ is the spin cumulant;
$\rho_I(\epsilon)$ represents the RKKY density of states, which is obtained from $\rho_I(\epsilon)=\sum_{\bm{q}} \delta (\epsilon - I_{\bm{q} }) $;
{the subscript ``loc" means that quantity is local 
at the site of an $f$-moment, and $m_{AF}$ is the AF magnetic moment.}
We consider a generic density of the conduction electrons ($\rho_{0}$) with a nonzero value at the Fermi energy.  
Finally, the lattice spin susceptibility at momentum $\bm{q}$ is calculated by 
\begin{equation}
\label{eqn:chi_latt}
    \chi^{a}
    (\bm{q}, i\omega_n) = \frac{1}{I_{\bm{q}} + M^{a}(i\omega_n)} \, .
\end{equation}

\noindent {\bf{Mutual information.~~}}
Mutual information measures the information between two subsystems.
Consider a subspace $A\times B \subset M$ where $M$ is the entire manifold of the parameter space of the system, the mutual information of a mixed state $\hat{\rho}_{AB}$ on this space is defined by~\cite{nielsen2010quantum}
\begin{align}
    {{\rm MI}}(A;B):=& S(\hat{\rho}_{A})+S(\hat{\rho}_{B}) - S(\hat{\rho}_{AB}) \label{eqn:MI_def1}\\
    \equiv & \tr\left( \hat{\rho}_{AB} (\log(\hat{\rho}_{AB}) - \log(\hat{\rho}_A \otimes \hat{\rho}_B)) \right) \label{eqn:MI_def2}
\end{align}
where $\hat{\rho}_A=\tr_B\hat{\rho}_{AB}$, $\hat{\rho}_B=\tr_A\hat{\rho}_{AB}$ and $S(\hat{\rho}) = -\tr(\hat{\rho}\log\hat{\rho})$ is the von Neumann entropy. {In practice, we use the natural logarithm with base $e$. 
Note that Eq.~\ref{eqn:MI_def2} is generally applicable for nonzero temperatures as well.
The trace is carried out in terms of the local Hilbert space, with each local state contributing a thermally averaged factor. In Fig.~\ref{fig:MI}, we present data for the three lowest temperatures.  The data overlap among these temperatures indicates that the temperatures are sufficiently low to capture the behavior of the ground state at zero temperature.}

\noindent {\bf{Quantum Fisher Information.}~~}
To pave the way for the calculations presented in this work, we motivate how summing over all the covariance of site pairs leads to the QFI, and elaborate on how the QFI can witness the multipartite entanglement and be measured in experiments.
{Specifically, we consider the $z$-component of the $f$-spin operators of the Kondo lattice.}

The clearest case arises in a pure state, for which the QFI is defined as the variance function 
{$F_Q=4\var({\widehat S^z})\equiv 4(\langle (\widehat S^z)^2 \rangle - \langle {\widehat S^z} \rangle^2) $}
{(see Supplementary Note 1). }
Consider a lattice comprising $N$ sites, and the operator 
{${\widehat S^z}=\sum_{i=1}^N {\widehat S^z}_i$}
where the operator 
{$\widehat S^z_i$}
acts only on site $i$. We find 
{(as derived in Supplementary Note 1) }
that the QFI is written as follows:
\begin{equation}
\label{eqn:Drho_OO}
    \R{\frac{F_{Q}}{4} = \sum_{i=1}^N \var({\widehat S^z}_i) + \sum_{i,j\neq i} \tr \left( \left(\hat{\rho}_{ij} - \hat{\rho}_i \otimes \hat{\rho}_j\right)  \widehat{S}^z_i \otimes \widehat{S}^z_j \right)}
\end{equation}
where $\hat{\rho}_i=\tr_{j\neq i}\hat{\rho}$ and $\hat{\rho}_{ij}=\tr_{l\neq i,j}\hat{\rho}$. The second term in Eq.~(\ref{eqn:Drho_OO}) measures the difference between $\hat{\rho}_{ij}$ and $\hat{\rho}_i \otimes \hat{\rho}_j$ which is non-vanishing if the states on site $i$ and $j$ are entangled. 
We recognize that the mutual information captures the difference between $\R{\ln}(\hat{\rho}_{ij})$ and $\R{\ln}(\hat{\rho}_i \otimes \hat{\rho}_j)$ 
[See Eq.~(\ref{eqn:MI_def2})].
Thus, the QFI measures the sum of pairwise entanglement between all pairs of sites, which contains the information of multipartite entanglement.

For calculations at nonzero temperatures, we consider the case of a mixed state $\hat{\rho}$.
The QFI is defined for the Hermitian operator 
{$\widehat{ S}^z$}
with the spectral decomposition $\hat{\rho}=\sum_{\lambda_n} \lambda_n |\lambda_n\rangle \langle\lambda_n|$, it is expressed as~\cite{hauke2016measuring,liu2020quantum}
\begin{equation}
\label{eqn:QFI_def}
    \R{F_{Q} = 2\sum_{nm}\frac{(\lambda_n-\lambda_m)^2}{\lambda_n+\lambda_m} S^z_{nm} S^z_{mn}.}
\end{equation}
where 
{$S^z_{nm} = \langle\lambda_n|{\widehat{S}^z}|\lambda_m\rangle$.}
Further details about the QFI are introduced in
{Supplementary Note 1.}
The QFI of mixed states at temperature $T$ 
is determined by the susceptibility of operator 
\R{$\widehat{S}^z$ }
[Ref.~\citenum{hauke2016measuring}] via
\begin{equation}
\label{eqn:QFI_sus}
    \R{F_{Q} = \frac{2}{\pi} \int_{-\infty}^\infty \tanh{\frac{ \omega}{2T}} \chi''
    (\omega) d\omega } \, ,
\end{equation}
where 
{$\chi''
(\omega)$ }
is the imaginary part of the susceptibility 
{$\chi
(\omega)= i\int^\infty_0 dt e^{i\omega t} \tr\left( \hat{\rho} [{\widehat{S}^z}(t),{\widehat{S}^z}(0)] \right)$.}

\noindent {\bf{QFI bounds from the 
normalized QFI.}~~~}
Assume ${\widehat{\cal O}}=\sum_{i=1}^N {\widehat{\cal O}}_i$ and a mixed state of dimension $N$ where ${\widehat{\cal O}}_i$ acts on the $i$-th particle (basis vector). 
If the ground state can be written as the product of some $k$-partite entangled states with $k\leq m$, the QFI density $f_Q=F_Q/N$ is bounded above by:
$f_Q \leq m (h_{\text{max}}-h_{\text{min}})^2$,
where $h_{\text{max/min}}$ is the maximal/minimal eigenvalue of the operator ${\widehat{\cal O}}_i$ 
(for a version of the proof, see the SI, Sec.~I.C). 
Consequently, if
\begin{align}
\label{eqn:upperbound}
    f_{Q} > m (h_{\text{max}}-h_{\text{min}})^2,
\end{align}
it witnesses the existence of at least $m+1$-partite entangled states in the system~\cite{hyllus2012fisher}. 
In the remainder of this work, we focus on spin-$1/2$ systems. {Here, we consider the $f$-spin operator of a Kondo lattice (an Anderson lattice at sufficiently large $U$, and the maximal and minimal eigenvalues are $h_\text{max}=1/2$ and $h_\text{min}=-1/2$. Then}
the normalized QFI density, nQFI, is $f_Q/(h_{\text{max}}-h_{\text{min}})^2 = f_Q$.


\vskip 0.5 cm
\noindent{\bf\large Data availability}
\\
{The data that support the findings of this study are either presented in the manuscript or available at }\href{https://doi.org/10.5281/zenodo.14814713}{https://doi.org/10.5281/zenodo.14814713}.

\vskip 0.5 cm
\noindent{\bf\large Code availability}
\\
The computer codes that were used to generate the data that support the findings of this study are available from the corresponding author upon request.

\medskip
\medskip
\noindent $^\dagger$ These authors contributed equally.

\bibliographystyle{naturemagallauthors}

\bibliography{reference}

\clearpage

\noindent{\bf Acknowledgment:}~~
We thank Gabriel Aeppli,
Fakher Assaad, Matthew Foster, 
Pontus Laurell, Matteo Mitrano, Han Pu, Allen Scheie, Oliver Stockert and Alan Tennant
for useful discussions. Work at Rice has primarily been supported by the 
the NSF Grant No.\ DMR-2220603
(Y.F.),
the AFOSR
Grant No.\ FA9550-21-1-0356 (M.M.,Y.W.)
the Robert A. Welch Foundation Grant No.\ C-1411 (L.C.), 
and 
the Vannevar Bush Faculty Fellowship ONR-VB N00014-23-1-2870 (Q.S.). 
The
majority of the computational calculations have been performed on the Shared University Grid
at Rice funded by NSF under Grant EIA-0216467, a partnership between Rice University, Sun
Microsystems, and Sigma Solutions, Inc., the Big-Data Private-Cloud Research Cyberinfrastructure
MRI-award funded by NSF under Grant No. CNS-1338099, and the Extreme Science and
Engineering Discovery Environment (XSEDE) by NSF under Grant No. DMR170109. H.H. acknowledges
the support of the European Research Council (ERC) under the European
Union's
Horizon 2020 research and innovation program (Grant Agreement No.\ 101020833). Work in
Vienna was supported by the Austrian Science Fund (projects I 5868-N - FOR 5249 - QUAST,
SFB F 86, Q-M$\&$S, and 10.55776/COE1, quantA) 
and the ERC (Advanced Grant CorMeTop, No.\ 101055088).
Several of us
(Y.F.,\,M.M.,\,Y.W.,\,L.C.,\,S.P.,\,Q.S.)
acknowledge the hospitality of the Kavli Institute for Theoretical Physics, 
supported in part by the National Science Foundation under Grant No. NSF PHY1748958, 
during the program ``A Quantum Universe in a Crystal: Symmetry and Topology across the Correlation Spectrum".
Q.S. and S.P. acknowledge the hospitality of the Aspen Center for Physics, 
which is supported by NSF grant No. PHY-2210452.

\vspace{0.2cm}
\noindent{\bf Author contributions}
\\
Q.S. conceived the research. Y.F., M.M., Y.W., L.C., H.H. and Q.S. carried out model studies. Y.F., Y.W., S.P. and Q.S. contributed to the comparison of our calculated
QFI with experimental measurements.
Y.F., M.M., Y.W., L.C. and Q.S. wrote the manuscript, with inputs from all authors.

\vspace{0.2cm}
\noindent{\bf Competing 
 interests}\\
The authors declare no competing 
 interests.
 
 \vspace{0.2cm}
 \noindent{\bf Additional information}\\
Correspondence and requests for materials should be addressed to 
Q.S. (qmsi@rice.edu).

\clearpage

\clearpage

\clearpage

\clearpage
\clearpage

\newpage

\onecolumngrid 
\begin{center}
\textbf{\large Supplementary information for: Amplified multipartite entanglement witnessed in a quantum critical metal}
\end{center}

\setcounter{secnumdepth}{2} 
\setcounter{equation}{0}
\setcounter{figure}{0}
\setcounter{table}{0}
\renewcommand{\theequation}{S\arabic{equation}}
\renewcommand{\thefigure}{S\arabic{figure}}

\setcounter{secnumdepth}{3}

\section*{\label{app:QFI}Supplementary Note 1: Introduction to quantum Fisher information}
Because the quantum Fisher information is not widely considered in the condensed matter community, we briefly introduce it here in connection to the present work. {For comprehensive reviews, we refer the readers to Refs.~\citenum{pezze2018quantum,liu2020quantum}. }

Entanglement is usually hard to detect experimentally.
Different from entanglement entropy and mutual information, correlations are much easier to measure.
In fact, certain combination of correlation functions can be arranged to witness the entanglement. The first example of such a witness is Bell's inequality, where a violation of the inequality implicates the existence of entanglement.
Bell's inequality, though, is designed to distinguish bipartite entangled spin states from states without any entanglement, i.e. separable states. To witness any multipartite entangled states, we rely on the quantum Fisher information, which turns out to be measurable in experiments{~\cite{hauke2016measuring}}.

Quantum Fisher information was first proposed in quantum metrology as the quantum version of Fisher information in the classical estimation theory in statistics~\cite{helstrom1969quantum,holevo2011probabilistic,kay1993fundamentals,pezze2018quantum}. It is the upper bound of the classical Fisher information in quantum mechanical systems~\cite{braunstein1994statistical,braunstein1996generalized} 
Therefore, it serves as the quantum Cram\'{e}r-Rao lower bound~\cite{kay1993fundamentals,pezze2018quantum} of the variance of the unbiased estimation of parameter $\theta$, $\var(\hat\theta)\geq F^{-1}_Q$. It has been found that entanglement can increase the precision of estimations beyond the classical limit~\cite{petz2011introduction,liu2020quantum,pezze2018quantum}.

Conversely, the quantum Fisher information can also detect the amount of entanglement. {As we will explain further below, this type of quantum Fisher information is associated with unitary processes and are defined for operators instead of parameters.}

In the following subsection~A, we will first review the definition of quantum Fisher information and derive the definition Eq.~(10) that we used in the main text. In a next subsection~C, we present a proof of the general upper bound of quantum Fisher information for generic operators~\cite{pezze2018quantum}.

\subsection{\label{app:QFI_def}The definition of quantum Fisher information}
To motivate the quantum Fisher information~\cite{liu2020quantum,braunstein1994statistical}, 
we consider a process where the density matrix evolves as follows
\begin{equation}
    \partial_a \hat{\rho} = \frac12 (\hat{\rho} \widehat{L}_a+\widehat{L}_a\hat{\rho}) \, .
\end{equation}
Here, $\partial_a=\partial/\partial\theta_a$, with $\theta_a$
parametrizes this process, and $\hat{\rho}$ is the density matrix.
This equation also defines $\widehat{L}_a^\dagger=\widehat{L}_a$ as the symmetric logarithmic derivative (SLD) operator of the density matrix.
{Associated with this process, the quantum Fisher information is defined as}~\cite{liu2020quantum,braunstein1994statistical}
\begin{equation}
    F_{Q} := \tr (\hat{\rho} \widehat{L}_a^2) \, . 
\end{equation}

For a pure state $|\psi\rangle$, $\hat{\rho}^2=\hat{\rho}$. Thus, $\widehat L_a = 2 \partial_a \hat{\rho}$ and the quantum Fisher information is{~\cite{liu2020quantum}}
\begin{align}
    F_Q(|\psi\rangle)  &=  4(\langle \partial_\theta\psi|\partial_\theta\psi \rangle - |\langle \psi|\partial_\theta \psi \rangle|^2) \, .
\end{align}
If we consider the unitary process $|\psi\rangle \rightarrow e^{-i \theta \widehat{\mathcal O}}|\psi\rangle$, the quantum Fisher information of pure state $|\psi\rangle$ is
\begin{equation}
    F_Q(|\psi\rangle) = 4\var (\widehat{\mathcal O} ) \, .
\end{equation}

For mixed states, we can use a spectral decomposition of $\hat{\rho}=\sum_i \lambda_n |\lambda_n\rangle\langle\lambda_n|$ to derive the following expression{~\cite{liu2020quantum}}
\begin{equation}
\label{eqn:QFI_def_par}
    F_{Q}= \sum_n \frac{(\partial_{a}\lambda_n)^2}{\lambda_n} + 2\sum_{nm} \frac{(\lambda_n-\lambda_m)^2}{\lambda_n+\lambda_m} \big|\langle \lambda_n| \partial_{a}\lambda_m \rangle \big|^2 \, .
\end{equation}
The first term vanishes when the state evolves unitarily since the eigenvalues $\lambda_n$ do not change.
Thus, the second term is the main focus. 
It describes the fluctuations of the states with respect to the parameter changes.

For the unitary process $|\lambda_n\rangle \rightarrow e^{-i\theta_a\widehat{\cal O}_a}|\lambda_n\rangle$, or equivalently $\hat{\rho} \rightarrow e^{-i\theta_a\widehat{\cal O}_a} \hat{\rho} e^{i\theta_a\widehat{\cal O}_a}$, using the evolution of wavefunction or the density matrix
{
\begin{align}
    i\partial_a |\lambda_n\rangle = \widehat{\cal O}_a |\psi \rangle \, , \qquad
    i\partial_a \hat{\rho} = [\widehat{\cal O}_a, \hat{\rho}] \, ,
\end{align}
}
we get the following simplified definition of quantum Fisher information is Eq.~(10) of the Methods section, which we reproduce here~\cite{liu2020quantum}:
\begin{equation}
\label{eqn:QFI_def_app}
    F_{Q} = 2\sum_{nm}\frac{(\lambda_n-\lambda_m)^2}{\lambda_n+\lambda_m} {\mathcal O}_{nm} {\mathcal O}_{mn}.
\end{equation}
Instead of taking the derivative of parameters, this definition only rely on the operators. 
Note, this quantum Fisher information corresponds to the particular unitary process described above. The process can be defined for any Hermitian operator $\widehat{\cal O}$. 
Typical combinations of parameter and operator in such unitary processes include position and momentum operator, angle and angular momentum operator, time and the Hamiltonian operator.

QFI is an additive and convex function of the density matrix given the operator ${\widehat{\cal O}}$~\cite{liu2020quantum,toth2015evaluating,pezze2018quantum}, i.e. 
\begin{align}
    F_{Q} \left(\sum_\alpha p_\alpha \hat{\rho}_\alpha \right) &\leq \sum_\alpha p_\alpha F_{Q} (\hat{\rho}_\alpha) \leq \max_{|\psi \rangle} F_{Q}(|\psi\rangle)
    \label{eqn:convex}
\end{align}
where $\sum_\alpha p_\alpha=1$ is a partition and the maximal value of $F$ is always reached by a pure state as the last inequality shows.

\subsection{\label{app:QFI_cov} Local operators detect multipartite entanglement}
QFI defined by Eq.~(\ref{eqn:QFI_def_app}) reduces to the variance of operator \R{$\widehat{\cal O}$} for a pure state, 
\begin{align}
    \frac14 F_{Q}(|\psi\rangle) = \var (\widehat{\cal O}) = \langle \psi | \widehat{\cal O}^2|\psi\rangle - \langle \psi | \widehat{\cal O}|\psi\rangle^2 \, .
\end{align}
If we select the operator as a sum of $N$ independent local operators $\widehat{\cal O}_i$, each exclusively acting on a given site, i.e. $\widehat{\cal O} = \sum_i^N \widehat{\cal O}_i$, the variance take the following form:
\begin{align}
    \var (\widehat{\cal O}) &= \sum_{i=1}^N \langle \widehat{\cal O}_i^2 \rangle - \langle \widehat{\cal O}_i \rangle^2 + \sum_j \sum_{i\neq j} \langle \widehat{\cal O}_i \widehat{\cal O}_j \rangle - \langle \widehat{\cal O}_i \rangle \langle \widehat{\cal O}_j \rangle \nonumber\\
    &=\sum_i \var_i + \sum_j \sum_{i\neq j} \text{Cov}_{ij} \, ,
    \label{eqn:var_O_app}
\end{align}
where the diagonal term $\var_i = \langle \widehat{\cal O}_i^2 \rangle - \langle \widehat{\cal O}_i \rangle^2$ is the variance of each local operator and the off-diagonal term $\text{Cov}_{ij}=\langle \widehat{\cal O}_i \widehat{\cal O}_j \rangle - \langle \widehat{\cal O}_i \rangle \langle \widehat{\cal O}_j \rangle $ is the covariance. The covariance detects the entanglement between sites $i$ and $j$ since it can be written as
\begin{align}
    \text{Cov}_{ij} &= \sum_{i\neq j} \tr \left( \hat{\rho} \widehat{\cal O}_i \widehat{\cal O}_j \right) - \tr \left( \hat{\rho} \widehat{\cal O}_i \right) \tr \left( \hat{\rho} \widehat{\cal O}_j \right) \nonumber \\
    &= \tr \left( \hat{\rho}_{ij} (\widehat{\cal O}_i \otimes \widehat{\cal O}_j ) \right) - \tr \left( (\hat{\rho}_i \widehat{\cal O}_i \otimes \hat{\rho}_j \widehat{\cal O}_j) \right) \nonumber \\
    &= \tr \left( \left(\hat{\rho}_{ij} - \hat{\rho}_i \otimes \hat{\rho}_j\right)  \widehat{\cal O}_i \otimes \widehat{\cal O}_j \right) \, ,
    \label{eqn:cov_app}
\end{align}
where $\hat{\rho}=|\psi\rangle\langle\psi|$, $\hat{\rho}_i=\tr_{j\neq i}\hat{\rho}$ and $\hat{\rho}_{ij}=\tr_{l\neq i,j}\hat{\rho}$.
The second equality in Eq.~(\ref{eqn:cov_app}) utilizes the fact $\tr A\otimes B = \tr A \tr B$.
Eq.~(\ref{eqn:cov_app}) shows that the covariance of two local operators at sites $i$ and $j$ is a measure of $\hat{\rho}_{ij} - \hat{\rho}_i \otimes \hat{\rho}_j$.
Therefore, this term measures the pairwise entanglement between the two sites.
\R{When the operators do not commute, e.g. for operators that anti-commute, the framework does not apply.}

Combining Eqs.~(\ref{eqn:QFI_def_app}) and~(\ref{eqn:cov_app}), we conclude that the variance of \R{$\widehat{\cal O}$}, which is the sum of the measurement of pairwise entanglement $\text{Cov}_{ij}$, contains the information of multipartite entanglement of the entire system. 
Eqs.~(\ref{eqn:var_O_app}) and (\ref{eqn:cov_app}) derive Eq.~(9) in the main text.

In discussing the properties of $\var(\widehat{\cal O})$ and $\text{Cov}_{ij}$, it is important to note two key aspects. First, $\var(\widehat{\cal O})\geq 0$ while $\text{Cov}_{ij}$ does not have such constraint and can be negative.
Secondly, if all local operators $\widehat{\cal O}_i$ have identical eigenvalues, these quantities are bounded by
\begin{align}
    \var(\widehat{\cal O}) &\leq N^2(h_{\text{max}}-h_{\text{min}})^2/4 \label{eqn:inequality1}\\
    \var_i &\leq (h_{\text{max}}-h_{\text{min}})^2/4 \label{eqn:inequality2}\\
    |\text{Cov}_{ij}| &\leq |h_{\text{max}}-h_{\text{min}}|\times \nonumber \\
    &\qquad \max \{ |h_{\text{max}}|, |h_{\text{min}}|, |h_{\text{max}}+h_{\text{min}}| \}  \, ,
    \label{eqn:inequality3}
\end{align}
where $h_{\text{max/min}}$ is the maximal/minimal eigenvalue of the operators $\widehat{\cal O}_i$.
The proof of Eq.~(\ref{eqn:inequality1}) and Eq.~(\ref{eqn:inequality2}) is shown in the next subsection~\ref{app:QFI_bound}. The bound in Eq.~(\ref{eqn:inequality3}) follows that $\langle \widehat{\cal O}_i\widehat{\cal O}_j\rangle \leq \max\{h_{\text{max}}^2,h_{\text{min}}^2\}$ and $ \langle \widehat{\cal O}_i\rangle\langle\widehat{\cal O}_j\rangle \geq \min\{h_{\text{max}}^2,h_{\text{min}}^2,h_{\text{max}}h_{\text{min}}\}$.

\subsection{\label{app:QFI_bound} The upper bound of quantum Fisher information}
Quantum Fisher information is a positive-semi-definite, additive and convex function of density matrix $\hat{\rho}$~\cite{liu2020quantum,toth2015evaluating,pezze2018quantum,braunstein1994statistical}
\begin{align}
    F_{Q} \left(\sum_\alpha p_\alpha \hat{\rho}_\alpha \right) &\leq \sum_\alpha p_\alpha F_{Q} (\hat{\rho}_\alpha) \leq \max_{|\psi \rangle} F_{Q}(|\psi\rangle) \, .
\end{align}
Therefore, the maximum value is realized by a pure state. Now we are interested in the question, what is the upper bound of the quantum Fisher information when the wavefunctions are restricted to be $m$-entangled, which means the density matrix can be decomposed into tensor products
\begin{equation}
    \hat{\rho} = \otimes_l \hat{\rho}_l \, ,
\end{equation}
where the dimension of each $\hat{\rho}_l$ is no greater than $m$.

For $N$ identical particles with operators $\widehat{\cal O}_j$, where $j=1\dots N$, the quantum Fisher information of a pure state at zero temperature is the variance
\begin{equation}
    F_{Q} = 4\left( \langle \widehat{\cal O}^2 \rangle-\langle \widehat{\cal O} \rangle^2 \right) \, ,
\end{equation}
where $\widehat{\cal O} = \sum_j \widehat{\cal O}_j$. We now show that for $m$-entangled states, the quantum Fisher information is bounded by 
\begin{equation}
    F_{Q} \leq m N(h_{\text{max}}-h_{\text{min}})^2 \, ,
\end{equation}
where $h_{\text{max/min}}$ is the maximal/minimal eigenvalue of the operators $\widehat{\cal O}_j$.

We present one proof of this bound of general operators, which is in agreement with Ref.~\citenum{pezze2018quantum}.

\begin{proof}
    We first consider the case $N/m \in \mathbb Z$. Then the $N$ particles can be divided into $N/m$ copies of $m$-entangled states. Therefore,
    \begin{align}
        F_{Q} &= 4\frac Nm \left(\langle \left( \sum_{j=1}^{m}\widehat{\cal O}_j \right)^2 \rangle-\langle \sum_{j=1}^{m}\widehat{\cal O}_j \rangle^2 \right) \nonumber \\
        & \leq 4\frac Nm~m^2 \left(\frac{h_{\text{max}}-h_{\text{min}}}{2}\right)^2 \nonumber \\
        & = Nm (h_{\text{max}}-h_{\text{min}})^2 \, ,
        \label{eqn:der_QFIbound}
    \end{align}
    where the inequality comes from the maximal variance of the operator $\widehat{\cal O}'=\sum_{j=1}^m\widehat{\cal O}_j$.
    To see this, consider the spectral decomposition
    \begin{align}
        &\langle \widehat{\cal O}'^2 \rangle - \langle \widehat{\cal O}' \rangle^2 = \sum_n p_n h_n'^2 - (\sum_n p_n h_n')^2 \nonumber\\
        &\leq p h_{\text{max}}'^2 + (1-p) h_{\text{min}}'^2 - (p h_{\text{max}}'+(1-p) h_{\text{min}}')^2 \nonumber\\
        &= -(h_{\text{max}}'-h_{\text{min}}')^2 (p^2-p) \nonumber\\
        &\leq \left(\frac{h_{\text{max}}'-h_{\text{min}}'}{2}\right)^2 \nonumber \\
        &\leq m^2 \left(\frac{ h_{\text{max}}- h_{\text{min}}}{2}\right)^2 \, ,
        \label{eqn:der_QFIbound_part2}
    \end{align}
    where $h'_n$ is the $n$-th eigenvalue of of $\widehat{\cal O}'$, $h'_{\text{max/min}}$ is the maximal/minimal eigenvalue of $\widehat{\cal O}'$, and $\sum_n p_n=1$ is a partition. The maximal variance is realized when only $h'_{\text{max/min}}$ are involved. The first and second inequality in Eq.~(\ref{eqn:der_QFIbound_part2}) derive the maximal variance of a general operator and the last inequality is a consequence of the fact $mh_{\text{min}} \leq h'_{\text{min}}\leq h'_{\text{max}}\leq mh_{\text{max}}$. 
    Therefore, applying Eq.~(\ref{eqn:der_QFIbound_part2}) to $\widehat{\cal O}'$ leads to Eq.~(\ref{eqn:der_QFIbound}).
    
    When $N/m\not\in \mathbb Z$, as a consequence of the convexity property~\cite{hyllus2012fisher}, the quantum Fisher information is maximized when the $[\frac Nm]$ sectors of $m$-entangled states and the residual sector of $(N-[\frac Nm]m)$-entangled states are all maximally entangled. This leads to
    \begin{equation}
    F_{Q} \leq \bigg( \big[\frac{N}{m} \big]m^2+r^2 \bigg) (h_{\text{max}}-h_{\text{min}})^2 \, ,
    \end{equation}
    where $\big[\frac{N}{m} \big]$ is the integer part of $N/m$ and $r=N-\big[\frac{N}{m} \big]$. In the case of $N\gg m$, this bound is
    \begin{align}
    F_{Q} &\leq \bigg( Nm+O(m^2) \bigg) (h_{\text{max}}-h_{\text{min}})^2 \nonumber \\
    &\xrightarrow[]{N\rightarrow \infty} Nm (h_{\text{max}}-h_{\text{min}})^2 \, .
    \end{align}
\end{proof}
This conclusion works for generic operators. We check that it agrees with spin-$\frac12$ chain where the bound is saturated by Greenberger–Horne–Zeilinger (GHZ) states~\cite{hyllus2012fisher}.

\section*{\label{app:two_tangle}Supplementary Note 2: Two-tangle}

One-tangle and two-tangle 
were proposed to be 
measures of quantum entanglement in qubits~\cite{PhysRevLett.80.2245}. 
One-tangle is defined as 
\begin{equation}
    \tau^{(1)} = 4\det{\hat{\rho}_1} \, ,
\end{equation}
where $\hat{\rho}_1$ is the reduced density matrix that  traces out all qubits except the target one. $\tau^{(1)} \in [0,1]$ measures the entanglement between the target qubit and the remaining qubits. When $\tau^{(1)} =0$, the qubit is separable from other qubits; when $\tau^{(1)} =1$, they are maximally entangled.

Two-tangle is defined as
\begin{equation}
    \tau = |\max\{0,2\lambda_{\text{max}}-\tr R \}|^2 \, ,
\end{equation}
where $\lambda_{\text{max}}$ is the maximal eigenvalue of matrix $R = \sqrt{\sqrt{\hat{\rho}} \tilde{\hat{\rho}} \sqrt{\hat{\rho}}}$, where $\tilde{\hat{\rho}} = {\cal T} \hat{\rho} {\cal T}^{-1}$ is the density matrix after spin-flipping. ${\cal T}=\sigma_y\otimes \sigma_y {\cal K}$ for spin-$1/2$ systems and $\cal K$ is the charge conjugation operator. 
Two-tangle $\tau\in[0,1]$;
it describes the entanglement between two quibits. 
When $\tau =0$ the two qubits are separable; when $\tau =1$ they are maximally entangled.
Two-tangle is also closely related to the entanglement of formation $E_f$, which measures the amount of entanglement to prepare a state from completely separable states. The definition of $E_f$ and its relation to two-tangle are~\cite{PhysRevLett.80.2245}
\begin{align}
    E_f &= \inf \left\{ \sum_n p_n E_f(|\psi_n\rangle) \right\} \\
    E_f &= h\left( \frac{1+\sqrt{1-\tau(\hat{\rho})}}{2}\right) \, ,
\end{align}
where the infimum is taken over all possible decomposition of $\hat{\rho} = \sum_n p_n |\psi_n\rangle\langle\psi_n|$ and, for a pure state, $E_f(|\psi\rangle_{AB})=S(\hat{\rho}_A)=S(\hat{\rho}_B)$ is the entanglement entropy. Here $h(x) = -x\log x-(1-x)\log(1-x)$. 

Two-tangles satisfy the Coffman-Kundu-Wootters inequality~\cite{coffman2000distributed,PhysRevLett.96.220503}
\begin{equation}
    \tau_{AA_1}+\tau_{AA_2}+\dots+\tau_{AA_n} \leq \tau_{A(A_1A_2\dots A_n)} \, ,
\end{equation}
which means the sum of pairewise entanglement between $A$ and $A_i$ is no greater than the entanglement between $A$ and the remaining system.
This inequality quantifies the concept of quantum monogamy, meaning that if $A$ and $A_1$ are maximally entangled (i.e. $\tau_{AA_1}=\tau_{A(A_1A_2\dots A_n)}=1$), then $A$ can not entangle with other qubits (i.e. $\tau_{AA_2}=\dots=\tau_{AA_n}=0$).

\begin{figure}[b!]
    \centering
    \includegraphics[width=0.9\linewidth]{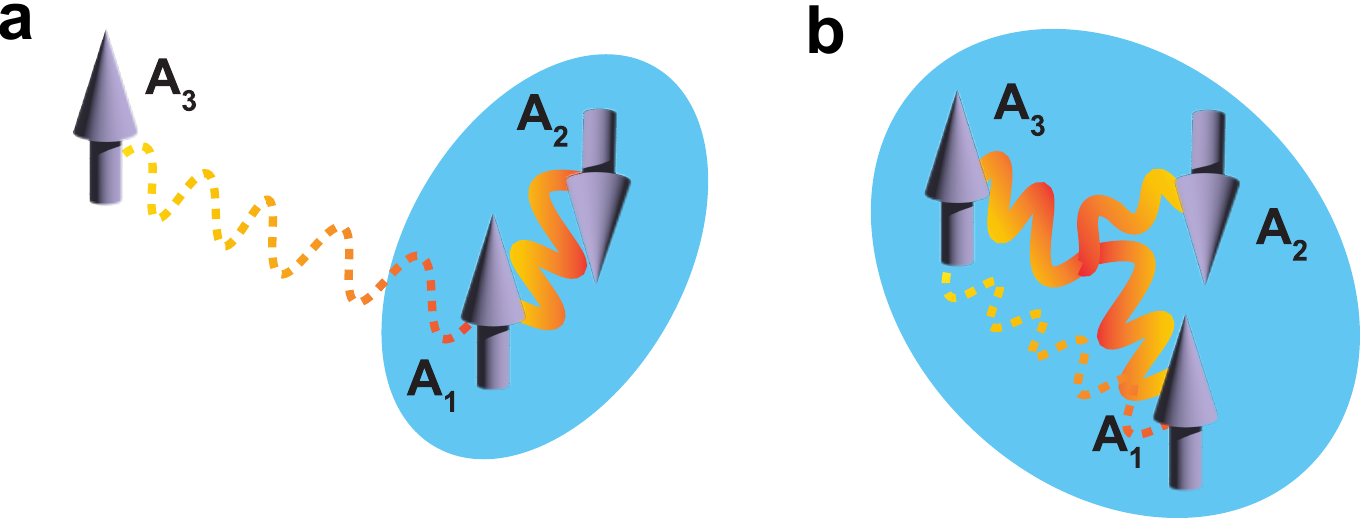}
    \caption{{{\bf a,} Schematics of quantum monogamy. If two qubits \R{(arrows on the figure)} $A_1$ and $A_2$ are maximally entangled \R{(represented by the solid wavy line)}, for example forming a Bell state, they cannot entangle with the third qubit $A_3$ \R{(with weak entanglement being represented by the dashed wavy line)}. {\bf b,} If three qubits are maximally entangled and the entanglement is in forms of three partite, then the bipartite entanglement is weak.  } }
    \label{fig:monogamy}
\end{figure}

{In Fig.~\ref{fig:monogamy} {\bf a}, we illustrate the quantum monogamy for three qubits. If qubits $A_1$ and $A_2$ are maximally entangled, e.g. forming a Bell state, then $\tau_{A_1A_2}=1$. Therefore, Coffman-Kundu-Wootters inequality shows 
$1=\tau_{A_1A_2} + \tau_{A_1A_3} \leq \tau_{A_1(A_2A_3)}=1$. Therefore, $\tau_{A_3(A_1A_2)}=\tau_{A_1A_3}=\tau_{A_2A_3}=0$.  }

{Fig.~\ref{fig:monogamy} {\bf b} illustrates maximally entangled three qubits with zero two tangle $\tau_{A_1A_2}=\tau_{A_1A_3}=\tau_{A_2A_3}=0$ and $\tau_{A_1(A_2A_3)}=\tau_{A_2(A_1A_3)}=\tau_{A_3(A_1A_2)}=1$. The residual tangle $\tau_{A_1(A_2A_3)} - \tau_{A_1A_2} - \tau_{A_1A_3}=1$ shows that the entanglement is distributed among the three qubits~\cite{coffman2000distributed}.   }

Two-tangle describes pairwise entanglement, but it fails to capture the multipartite entanglement. For example, the
GHZ state $|\psi\rangle = \frac{1}{\sqrt{2}}(|\uparrow\uparrow\uparrow\rangle + |\downarrow\downarrow\downarrow\rangle) $ is a maximally entangled state. Its two-tangle of each qubit pair vanishes, which indicates 
the absence of pairwise entanglement in this state. This is because the entanglement in the GHZ state is spread out over all the spins instead of pairing of two particular spins.

In previous literature, two-tangles have been used to study one-dimensional spin chains where the two-tangles can be expressed as functions of equal time correlation functions due to the spin parity and translation symmetries~\cite{amico2004dynamics}.

\begin{figure}[b!]
    \centering
    \includegraphics[width=0.8\linewidth]{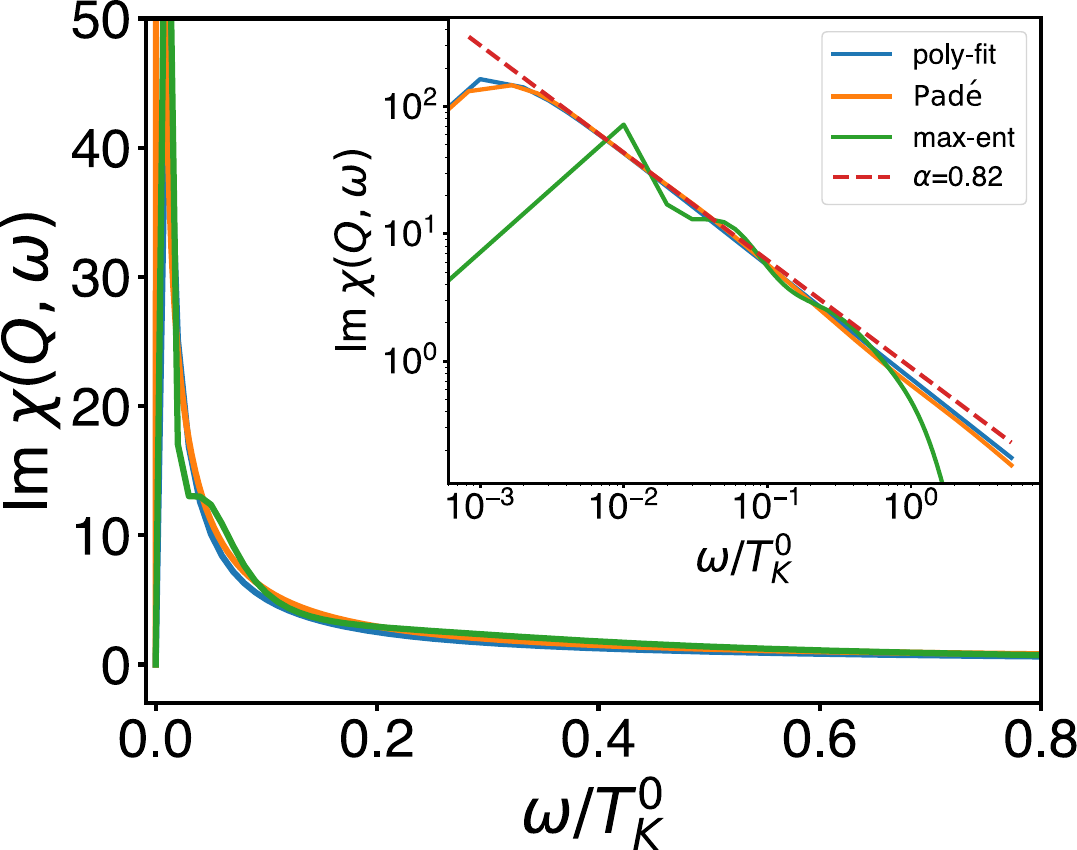}
    \caption{
    {\bf Comparison between the different methods of analytical continuation.~~} The blue, orange and 
    green curves correspond to $\R{\rm Im} \chi(\R{\mathbf Q}, \omega)$ at $T/T_K^0=2.5\times 10^{-3}$  obtained from the polynomial fitting, maximal entropy method and 
    Pad\'e decomposition, respectively. 
    \R{The inset displays the plot on a double logarithmic scale.}
    }
    \label{fig:ana-cont}
\end{figure}

For spin-$\frac12$ isotropic antiferromagnetic models without order, the two tangle is~\cite{amico2004dynamics}
\begin{equation}
    \tau_{0\mathbf{R}} = \left|\max \left\{ 0,-\frac12 \pm 2\langle \mathbf{S}_0\cdot \mathbf{S}_\mathbf{R}\rangle \right\}\right|^2 \, .
\end{equation}
It equals zero for $|\langle \mathbf{S}_0\cdot \mathbf{S}_\mathbf{R}\rangle| < 0.25$ for $\mathbf R \neq 0$ (note $\langle \mathbf{S}_0 \cdot \mathbf{S}_0\rangle\equiv 3/4$ has no information for entanglement).
The one-tangle for spin systems can be simplified as~\cite{amico2004dynamics}
\begin{equation}
    \tau^{(1)} = 1 - 4 \sum_{a=x,y,z} \langle S^a\rangle^2 \, .
\end{equation}
In the absent of order, $\tau^{(1)} = 1$ is maximized.
The sum of two-tangles $\sum_{\mathbf R\neq 0}\tau_{0{\mathbf R}}$ can be used to describe the amount of pairwise entanglement in the system while the residual tangle $\tau^{(1)}-\sum_{{\mathbf R}\neq 0} \tau_{0{\mathbf R}}$ can be used to describe the entanglement stored in other forms~\cite{amico2004dynamics}. In $\rm KYbSe_2$, which realizes the antiferromagnetic spin-$1/2$ Heisenberg model on a triangular lattice~\cite{scheie2023proximate}, the two-tangle was determined as a part of the entanglement characterization by inelastic neutron scattering measurements.

\section*{Supplementary Note 3: Analytical continuation}

In order to obtain the dynamical spin susceptibility in 
real frequency, we use three types of methods to 
perform an analytical continuation: the Pad\'e decomposition, maximal entropy and polynomial fitting. 
The polynomial fitting here refers to 
a fit
of 
$\chi(\R{\mathbf Q}, i\omega_n)$ in the imaginary frequency by $\frac{a}{(\omega_n+b)^c}$, which is followed by the substitution $i \omega_n\to \omega + i 0^+$ to obtain $\chi(\R{\mathbf Q}, \omega)$ in real frequency.

A comparison of the results at QCP with $T/T_K^0=2.5\times10^{-3}$ is displayed in Fig.~\ref{fig:ana-cont}, which shows a high degree of compatibility between the results derived from the different methods. 
The values of QFI obtained by these three methods are also comparable with each other. In the low temperature limit, we further 
use $f_Q \sim 4\chi(\R{\mathbf Q},\tau=0)=\frac{4}{\beta} \sum_{n} \chi (\R{\mathbf Q}, i\omega_n) $
to directly determine the QFI from the Monte Carlo data; the calculated value if comparable to those obtained from the aforementioned three methods.

\begin{figure}[t!]
    \centering
    \includegraphics[width=0.55\linewidth]{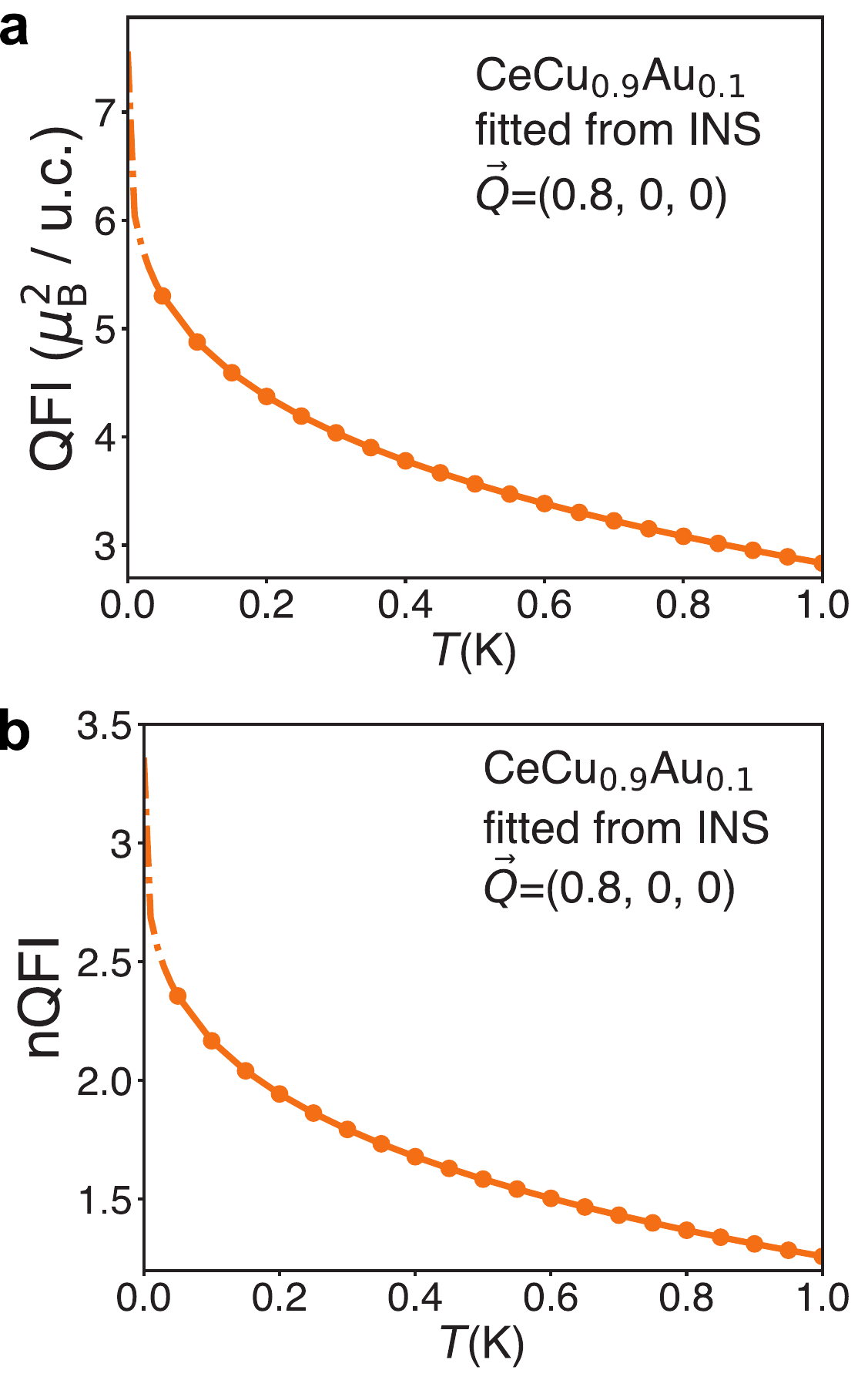}
    \caption{
    {\bf The (normalized) QFI density extracted from inelastic neutron scattering experiments in $\rm CeCu_{0.9}Au_{0.1}$.~~} 
    {\bf a,} The temperature dependence of the QFI density in
    CeCu$_{0.9}$Au$_{0.1}$, for the AF magnetization operator as probed by the 
    inelastic neutron scattering measurements,
    in unit of $\mu_B^2$ per unit cell.
    The solid curve 
    is fitted from the data 
    above the lowest temperature  measured in Ref.~\citenum{schroder2000onset}. 
    The dot-dashed curve 
    is an extrapolation 
    below the lowest measured temperature, based on the fitted expression Eq.~(\ref{scaling-Im-chi}). $\mathbf{Q}=(0.8,0,0)$ is among the peak wavevectors of the measured dynamical spin susceptibility. {\bf b,} The corresponding normalized QFI density (nQFI) of the AF spin operator, with $g=1.5$ (Ref.~\citenum{schroder2000onset}).
    }
    \label{fig:S2}
\end{figure}

\section*{Supplementary Note 4: Analysis of the quantum Fisher information in the heavy fermion quantum critical point of CeCu$_{5.9}$Au$_{0.1}$}

For heavy fermion metals, the spin spectral weight of the $f$-states is concentrated in a narrow energy range set by the Kondo temperature scale, while that of the conduction $c$ electron states has a broader distribution over the energy range of a bare conduction electron bandwidth. 
Since the inelastic neutron scattering spectrum is appreciable only in the energy range of the Kondo temperature scale, the spin spectrum we have calculated in the determination of the QFI, based on the $f$-spin operators, is a suitable quantity to compare with experimental measurements in heavy fermion metals.

CeCu$_{0.9}$Au$_{0.1}$ is a canonical heavy fermion metal that is located at a 
quantum critical point~\cite{schroder2000onset}.
Here, we utilize the data from the inelastic neutron scattering and 
magnetometry measurements \cite{schroder2000onset} to
extract the quantum Fisher information. The dynamical spin susceptibility has been parameterized as follows\cite{schroder2000onset}:
\begin{align}\label{scaling}
    \chi_{exp}(\hbar\omega,T,\mathbf{q})=\frac{c}{\Theta(\mathbf{q})+(-i\hbar\omega/a +k_{B} T)^{\alpha}} \, ,
\end{align}
where the exponent is
$\alpha \approx 0.75$.
The coefficient 
was fitted to be $a \approx 0.8$.
The generalized Curie constant $c$ 
captures the
slope in the plot of $1/\chi_{exp}(\hbar\omega=0,T,q) $ versus $ (k_{B}T)^{\alpha}$.
The neutron scattering and bulk
susceptibility 
data yield (Ref.\,\citenum{schroder2000onset}, Fig.\,3a) a slope in $1/\chi_{exp}(E=0,T,\mathbf{q})$ vs. $T^{0.75}$ 
of $\approx 0.085\, {\rm meV}\,\cdot\,\mu_{\rm B}^{-2}\,\cdot\,{\rm K}^{-0.75}$.
This implies $c^{-1} *\left(8.62 \times 10^{-2} \, \text{meV/K}\right)^{0.75} \approx 0.085 \,\text{meV} \cdot \mu_{\rm B}^{-2} \cdot \text{K}^{-0.75}$, leading to
$c\approx 1.9 \mu_{\rm B}^{2}\cdot {\rm meV}^{-0.25}$. 

At the AF wave vector $\mathbf{Q}$, the ``modified Weiss field" $\Theta(q)$ is small and, to a good approximation, can be taken to be zero (i.e., the system is very close to the QCP, which we consider here).  The imaginary part of the dynamical spin susceptibility obtained by taking the imaginary part of Eq.~(\ref{scaling}) then becomes:
\begin{align}
\label{scaling-Im-chi}
\chi''_{exp}(\hbar\omega,T,\mathbf{Q})=\frac{
c\,\sin[\alpha\arctan(\hbar\omega/ak_{B}T)]}
{ (\hbar^{2}\omega^2/a^{2} + k_{B}^2 T^2)^{\alpha/2}} \, .
\end{align}

The normalized quantum Fisher information density (nQFI) of the AF spin operator is determined from 
the imaginary part of the dynamical spin susceptibility at the AF wave vector 
$\mathbf{Q}$:
\begin{equation}
    f_{Q} = \frac{4}{\pi(g\mu_{B})^{2}} \int_{0}^\Lambda \tanh{\frac{ \hbar\omega}{2k_{B}T}} \chi_{exp}''(\hbar\omega,T,\mathbf{Q}) d(\hbar\omega) \, ,
\end{equation}
where the $g$ factor appropriate for the low-energy scaling regime is $g=1.5$ (Ref.~\citenum{schroder2000onset}).
The QFI density of the AF magnetization operator, given by $f_Q(g\mu_B)^2$, can be obtained directly from the experimental data, while determining the normalized QFI density of the spin operator,  $f_Q$, requires further knowledge of $g$.
By setting the frequency cutoff $\Lambda = 1.0\, {\rm meV}$, which is on the scale of the bare Kondo temperature, we extract the QFI density (labeled QFI on the plot) vs. $T$, which is shown in Fig.\,\ref{fig:S2}{\bf a} from the lowest measured temperature $T=0.05$\,K and above (solid curve). 
We can extrapolate the $f_Q$ to temperatures below the measured range by applying the analytical expression,
Eq.~(\ref{scaling-Im-chi}).
This extrapolation is illustrated by the dot-dashed curve in Fig.\,\ref{fig:S2}{\bf a}. 
The corresponding normalized QFI density of the AF spin operator is shown in Fig.\,\ref{fig:S2}{\bf b}. 
The nQFI density $f_Q$ at the lowest measured temperature is about $2.36$
[and it extrapolates to $\approx 3.3$ at $T=0$, as seen from the dot-dashed curve in Fig.\,\ref{fig:S2}{\bf b}].

\begin{figure}[p]
    \centering
    \includegraphics[width=\linewidth]{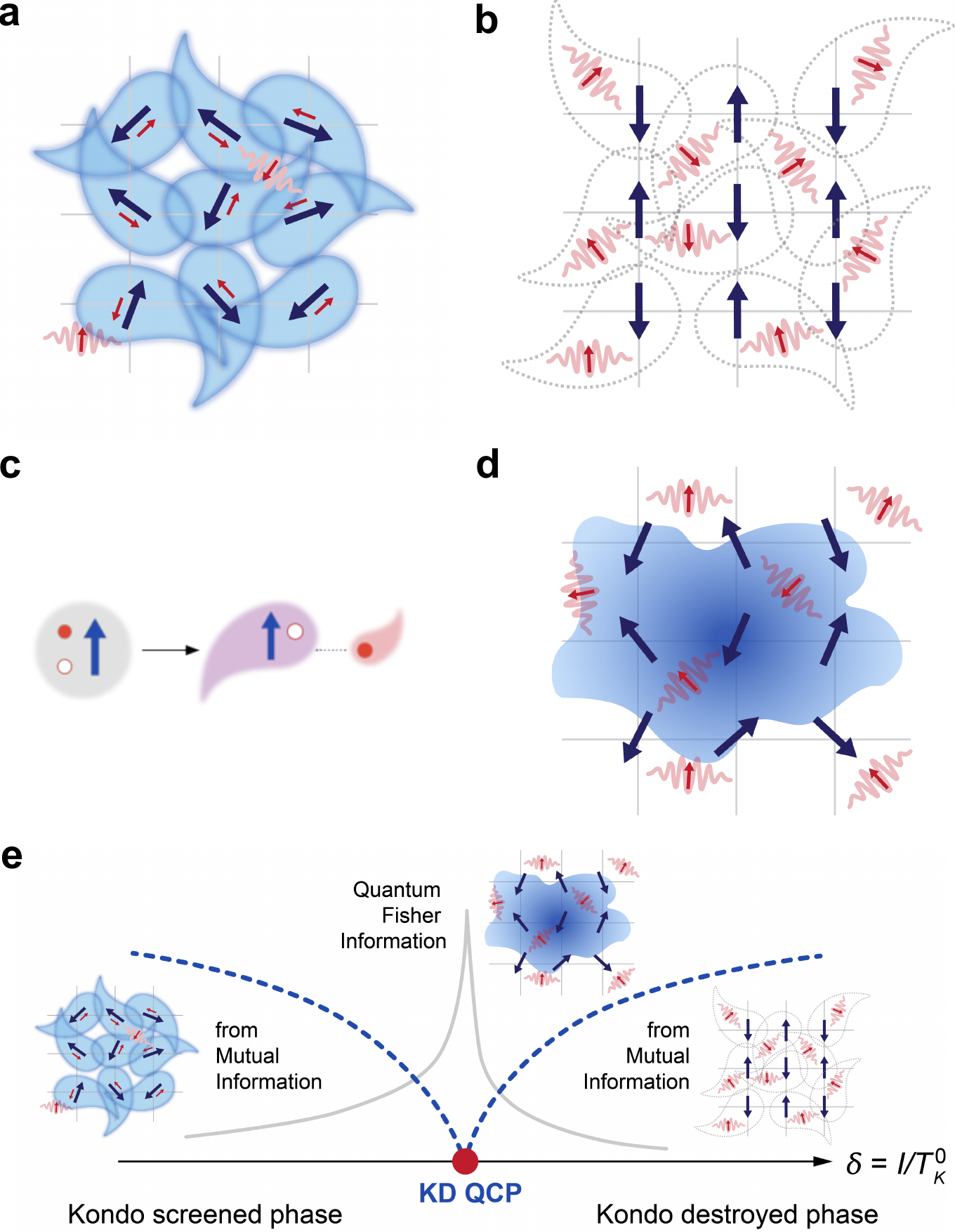}
\end{figure}
\begin{center}
    \captionof{figure}{
    \R{\bf 
    Quantum Fisher information characterizes loss of strange metal's quasiparticles.}
    {\bf a,} In the paramagnetic phase with Kondo screening,
    the saturated value of the calculated mutual information implicates that the $f$ moment and $c$ electrons form well-defined Kondo singlets.
    \R{Here the big and small arrows represent the $f$ and $c$ electron spins, respectively. A wavy line with the $c$-electron spin represents a free conduction electron. A blue droplet represents a Kondo singlet.}
    {\bf b,} In the antiferromagnetic phase,
    the calculated mutual information suggests suppressed Kondo singlets.
    {\bf c,} In the Kondo screened phase {\bf a}, the quasiparticle excitations (the composite between the $f$ moment and $c$ electron) develop out of the $f$ moment (left portion); this is enabled by the formation of the Kondo singlet in the ground state (left portion).
    \R{Here the solid/blank circles represent the $f$ electron/hole.}
     \R{Panels (a-c), as implicated by our mutual information calculation, are  adapted from Ref.~\citenum{hu2022quantumc}.}
    {\bf d,} In the quantum critical regime, the calculated QFI characterizes the entanglement between the $f$ moments, and this entanglement destabilizes the process of quasiparticle production and, hence, characterizes the loss of quasiparticles.
    \R{Here the 
    multipartite entanglement among 
    the spins are represented by the blue cloud on the figure.}
    \R{{\bf e} Schematics of how the QFI calculation characterizes strange metal's loss of quasiparticles in the regime associated with the Kondo-destruction quantum critical point (KD QCP).
    The dashed lines separate the quantum critical regime from the phases on the two sides and the gray line indicates the emergence of the 
    nQFI that is nontrivial (witnessing multipartite entanglement) in the quantum critical regime.}
    }
    \label{fig:Kondo_SM}
\end{center}
\clearpage

\section*{\label{app:Kondo}
Supplementary Note 5:
Entanglement implications for strange metallicity
}

{In this section, we extend on our discussion of the entanglement implications for the phase diagram of the Kondo lattice [Fig.~1 {\bf c}] and, in particular, the strange metallicity in the quantum critical regime.
We illustrate the behavior of the $f$ and $c$ electrons in the paramagnetic and antiferromagnetic phases and the quantum critical regime that is associated with the quantum critical point.
The control parameter of the phase diagram is the ratio of the RKKY coupling to the bare Kondo temperature $\delta=I/T_K^0$. When $\delta$ is small, it is in the Kondo screened phase. In the main text, we showed that the mutual information saturates to a maximal bipartite entanglement of a spin singlet. This implicates Kondo entanglement between the $f$ moments and conduction $c$ electrons, as illustrated in Fig.~\ref{fig:Kondo_SM} {\bf a}. For this phase, Fig.~\ref{fig:Kondo_SM} {\bf c} illustrates the quasiparticle excitation above the singlet ground state. The quasiparticle is a composite heavy fermion, and is in $1$-to-$1$ correspondence with the $f$ moments.
\R{Note that this correspondence applies for generic metallic cases, when the number of conduction electrons differs from that of the $f$-moments \cite{Oshikawa}.}}

When $\delta$ is large, the RKKY interaction dominates and the ground state is AF ordered.
The reduced mutual information (which vanishes in the large $\delta$ limit) implicates that the $f$ moments and conduction $c$ electrons have diminished entanglement, as illustrated in Fig.~\ref{fig:Kondo_SM} {\bf b}.

Our central result is that, in the quantum critical regime, the QFI of the spin operator for the $f$-moments witnesses multipartite entanglement among the $f$ moments. This implicates inter-$f$-moment entanglement of the RVB type [Fig.~\ref{fig:Kondo_SM} {\bf d}].
Through entanglement monogamy, this $f$-moment entanglement 
destabilizes the Kondo entanglement. Accordingly, 
it characterizes the lack of quasiparticle formation from the local moments.
\R{The overall picture is illustrated in Fig.~\ref{fig:Kondo_SM} {\bf e}, which shows how the multipartite entangelement, revealed by the QFI calculation, characterizes strange metal's loss of quasiparticles in the quantum critical regime.}

\R{ \section*{\label{app:parameter}Supplementary Note 6:
Further details of the EDMFT calculations} }

\R{In practice, we treat the bare density state of conduction electron as $\rho_c(\omega) = \frac{1}{2D}\Theta(|\omega-D|)$, where $D=1$ is the half bandwidth, to capture the nonzero density of states of the conduction electrons. Without a loss of generality, we set $U=0.25$, $V=0.4$, such that the effective Kondo temperature is sufficiently high so that our CTQMC calculation can reach sufficiently low temperatures to access an adequate dynamical range for quantum criticality (c.f., Fig.~3{\bf a} in the main text). The highest temperature shown in Fig.~2 of the main text corresponds to $\beta=200$.}

\R{In our calculations, we first perform a sufficiently large number of Monte Carlo samplings to ensure  that the bath was fully converged, followed by measurements that average over 200000 configurations. The errors in quantities computed in Matsubara frequencies are sufficiently small, with an average relative error of approximately $10^{-4}$. 
}

\end{document}